\documentclass[preprint]{aastex}
\usepackage{amsmath}
\usepackage{xspace}

\def\ie{i.e.\xspace}

\def\ltsima{$\; \buildrel < \over \sim \;$}
\def\simlt{\lower.5ex\hbox{\ltsima}}
\def\gtsima{$\; \buildrel > \over \sim \;$}
\def\simgt{\lower.5ex\hbox{\gtsima}}
\def\fesc{{$\langle f_{esc}\rangle$}\xspace}

\def\h2{H$_2$\xspace} 
\def\m{$^{-1}$\xspace}
\def\mm{$^{-2}$\xspace}
\def\mmm{$^{-3}$\xspace}

\received{.......}
\accepted{.......}
\slugcomment{to be submitted to ApJ }

\begin{document}
\tighten
\thispagestyle{empty}

\pagestyle{myheadings}
\markright{DRAFT: \today\hfill}

\def\placefig#1{#1}

\title{FEEDBACK FROM GALAXY FORMATION:   
PRODUCTION AND PHOTODISSOCIATION OF PRIMORDIAL \h2} 

\author{MASSIMO RICOTTI, NICKOLAY Y. GNEDIN AND J. MICHAEL SHULL$^1$ } 

\affil{
Center for Astrophysics and Space Astronomy \\
Department of Astrophysical and Planetary Sciences \\
University of Colorado, Campus Box 389, Boulder CO 80309 \\
E--mail: ricotti, gnedin, mshull@casa.colorado.edu \\ 
$^1$ also at JILA, University of Colorado and National
Institute of Standards and Technology }  
 
\begin{abstract}
 
  We use 1D radiative transfer simulations to study the evolution of
  H$_2$ gas-phase (H$^-$ catalyzed) formation and photo-dissociation
  regions in the primordial universe. We find a new {\em positive
    feedback} mechanism capable of producing shells of H$_2$ in the
  intergalactic medium, which are optically thick in some Lyman-Werner
  bands. While these shells exist, this feedback effect is important in reducing the H$_2$
  dissociating background flux and the size of photo-dissociation
  spheres around each luminous object. The maximum background opacity of the
  IGM in the H$_2$ Lyman-Werner bands is $\tau_{H_2} \approx 1-2$ for
  a relic molecular fraction
  $x_{H_2}=2 \times 10^{-6}$, about 6 times greater than found by
  \cite{HaimanAR:00}.  Therefore, the relic molecular hydrogen can
  decrease the photo-dissociation rate by about an order of magnitude.
  The problem is relevant to the formation of small primordial
  galaxies with masses $M_{DM} \la 10^8$ M$_\odot$, that rely on
  molecular hydrogen cooling to collapse.  Alternatively, the universe
  may have remained dark for several hundred million years after the
  birth of the first stars, until galaxies with virial temperature
  $T_{vir} \ga 10^4$ K formed.

\end{abstract}

\section{Introduction \label{sec:int}}

Many recent observations point to a flat geometry and a significant
cold dark matter (CDM) content. Recent measurements of the Cosmic
Microwave Background (CMB) anisotropies
\citep{deBernardis:00,Lange:00,Tegmark:00}, large-scale structure
\citep{Hamilton:00}, Ly$\alpha$ power spectrum \citep{Croft:99,
  McDonald:00}, and distances of high redshift SNe
\citep{Perlmutter:98,Riess:98,Garnavich:98} are all consistent with a
flat CDM model of the universe, in which about 1/3 the total energy
density of the universe is dark matter (DM) and 2/3 is ``vacuum
energy''. Big-Bang nucleosynthesis fits to the observed $D/H$
abundances suggest a small baryon density, with closure parameter
$\Omega_bh^2 \sim 0.02$ \citep{Burles:98}. Thus, the baryons are a
small fraction of the DM.  In the linear phase of structure
formation, the baryons simply trace the DM perturbations on large
scales.

In CDM cosmologies, small mass DM virialized objects form first.
Larger halos form later from the merger of smaller subunits
(hierarchical scenario).  Since the gas temperature prior to
reionization is about 100 K, the gas can collapse in objects with
M$_{DM} >10^4-10^5$ M$_\odot$. Small mass DM halos will then produce
the first luminous objects (Pop~III) if the baryons collapsed into the
DM potential wells and heated to the virial temperature can dissipate
their internal energy support. The lack of metals in the primordial
universe makes molecular hydrogen the only coolant below $T \sim 10^4$
K.  Therefore, unless some H$_2$ is formed, the baryons can not cool
in shallow DM potential wells with M$_{DM} \la 10^8 [(1+z)/10]^{-3/2}$
M$_\odot$, corresponding to a virial temperature $T_{vir} \la 10^4$ K.
The formation of H$_2$ occurs mainly through the chemical reaction,
\begin{equation}
H + H^- \rightarrow H_2 + e^- .
\end{equation}
Together with a similar reaction ($H_2^+ + H \rightarrow H_2 + H^+$),
this is the only possibility for virialized objects with $T_{vir} \la
10^4$ K to cool and eventually form stars \citep{Lepp:84}.  Despite
the fact that the physics is much simpler, theoretical models of the
formation and fate of Pop~III objects must confront the lack of
observational constraints and uncertainties on initial conditions. The
fate of the Pop~III objects is still unresolved or at least
controversial. The main complications of the model are the feedback
processes between luminous objects and the formation and destruction
of H$_2$ that can affect star formation on both local (interstellar
medium, ISM) and large (intergalactic medium, IGM) scales.

Molecular hydrogen is photo-dissociated by photons with energies between
$11.1$ and $13.6$ eV (Lyman-Werner bands) through the two step Solomon process
\citep{Stecher:67}.  Thus, the radiation from the first stars (or
quasars) could dissociate H$_2$ before the IGM is reionized ({\em
  negative feedback}). The H$^-$ formation process instead relies on
the abundance of free electrons.  A {\em positive feedback} is
therefore possible in a gas partially ionized by an X-ray background
or direct flux of Lyman continuum radiation.  Depending on the
relative importance of the positive/negative feedback, star formation
could be triggered or suppressed after the formation of the first
stars.

Understanding the star formation history at high redshift is important
for modeling the metal enrichment of the IGM. It may also influence
IGM reionization if, as pointed out in \cite{RicottiS:00}, the
contribution to the Lyman continuum (Lyc) emissivity in the IGM at
high redshift is dominated by low-mass galactic objects. The
contribution to the emissivity from these objects is important because
their spatial abundance is higher and because the fraction, \fesc, of
Lyc radiation that escapes from a low-mass galaxy halo is larger than
for more massive objects.

\cite{Tegmark:97} used a simple collapse criterion, that the gas
cooling time must be less than the Hubble time, to determine the
minimum H$_2$ fraction a virialized object must have in order to
collapse. This minimum mass is a decreasing function of the virialization
redshift and it is somewhat dependent on the cooling
function used. \cite{Tegmark:97} found a minimum mass of $M_{DM} \sim
10^6$ M$_\odot$ at $z \sim 30$ using the \cite{Hollenbach:79} molecular cooling
function. \cite{AbelAnninos:98} and \cite{Fuller:00} instead found a
minimum mass an order of magnitude smaller using \cite{Lepp:84} and
\cite{Galli:98} cooling functions respectively. Objects of this masses
can form around redshift $z \sim 30$ from $3\sigma$ perturbations of
the initial Gaussian density field. More realistic models
\citep{OmukaiN:98,Nakamura:99} confirm this basic result.

\cite{Abel:00} and \cite{BrommCL:99} used high-resolution numerical
simulations to study the formation of the first stars. \cite{Abel:00}
used 3D adaptive mesh refinement (AMR) simulation with dynamic range
about $2 \times 10^5$ to follow the collapse of the first objects starting from
cosmological initial conditions (standard CDM cosmology).  They found
an approximatively spherical protogalaxy of $10^6$ M$_\odot$ with a
collapsing core of $\sim 100$ M$_\odot$.
Similar results have been
found by \cite{BrommCL:99} using SPH simulations starting from ad hoc
initial conditions. For some initial conditions and larger halo masses
they could form disk protogalaxies with more than one star formation
region.  Unfortunately, without taking into account radiative transfer
effects and SNe explosions, it was not possible to determine the mass
or the initial mass function (IMF) of the first stars.  Local effects
of UV radiation on star formation have been studied with
semi-analytical models \citep{OmukaiN:99,Glover:00}. The main result
is that massive stars are effective in suppressing cooling in small
protogalaxies, but star formation can continue in larger galaxies. If
the ISM is clumpy, star formation is not suppressed in the denser
clumps.

The photo-dissociating UV radiation can affect the H$_2$ abundance
over large distances if the IGM is optically thin in the Lyman-Werner
bands.  \cite{Haiman:97} and \cite{HaimanAR:00} computed
self-consistently the rise in the dissociating UV background (UVB)
using the Press-Schechter formalism for sources of radiation. They
found that H$_2$ in the IGM had a negligible effect on the build-up of
the UVB because of its small optical depth $\tau_{H_2} \la 0.1$ and
because the photo-dissociation regions around the first luminous
objects overlap at an early stage. Therefore, the negative feedback of
the UVB suppresses star formation in small objects. However, if the
first objects are mini-quasars, the produced X-ray background is
strong enough to cancel out the UVB negative feedback, and
reionization could be produced by small objects.  Contrary to
\cite{HaimanAR:00} result, \cite{CiardiFA:00} found no negative
feedback from the UVB. They estimated the mean specific intensity of
the continuum background at $z=20$ in the $912-1120$ \AA\ Lyman-Werner
bands, $J_\nu \sim 10^{-27}$ erg s\m cm\mm Hz\m sr\m. This value is
four orders of magnitude smaller than the value found by
\cite{HaimanAR:00}.  Even if the photo-dissociation regions overlap by
$z =25$, only at $z \la 15$ does the UVB become the major source of
radiative feedback.  At higher redshifts, direct flux from neighboring
objects dominates the local photodissociation rate \citep{CiardiF:00}.
\cite{Machacek:00} included the feedback effect of the photo-dissociating
UV background in 3D AMR simulations of the formation and collapse of
primordial protogalaxies.  They used a box of 1 Mpc$^3$ comoving
volume, and resolution in the maximum refinement zones of about 1
pc comoving.  Their results confirm that a photo-dissociating
background flux of $F_{LW} \ga 5 \times 10^{-24}$ erg s\m cm\mm Hz\m
delays the cooling and star formation in objects with masses in the
range $10^5-10^7$ M$_\odot$.  They also provided an analytical
expression for the collapse mass threshold, for a range of UV fluxes.

The simple observation that star formation in our Galaxy can trigger
further star formation in a chain-like process \citep{McCray:87}
suggests that positive feedback effects could be important also at
high redshift.  \cite{HaimanRL:96} found that UV irradiation of a halo
can enhance the formation of H$_2$ and favor the collapse of the
protogalaxies.  \cite{Ferrara:98} estimated that the formation of
H$_2$ behind shocks produced during the blow-away process of Pop~III
can replenish the relic H$_2$ destroyed inside the photo-dissociation
regions around those objects. Finally, it has been noticed that
positive feedback is possible if the first objects emit enough X-rays
\citep{HaimanAR:00, Oh:00, Venkatesan:01}.

In this paper, we explore a new positive feedback effect that could be
important in regulating star formation in the first galaxies. Ahead of
the ionization front, a thick shell of several kpc of molecular
hydrogen (with abundance $x_{H_2}=n_{H_2}/n_H \la 10^{-4}$) forms because
of the enhanced electron fraction abundance in the transition region
between the H~II region and the neutral IGM. This shell, which we call
a {\em positive feedback region} (PFR), can be optically thick
in the H$_2$ Lyman-Werner bands, depending on the source turn-on
redshift $z_i$, luminosity, and escape fraction \fesc of ionizing
radiation.  We find that the photo-dissociation spheres around single
objects are smaller than in previous calculations where the effects of
the PFRs and H$_2$ self-shielding have been neglected. PFRs could also
be important in calculating the build-up of the dissociating
background. A self-consistent calculation of feedback effects on star
formation, including the new feedback process discussed in this study,
will be the subject of a subsequent paper. We will use a 3D
cosmological simulation with radiative transfer, in order to
understand what regulates the star formation at high redshift.

This paper is organized as follows. In \S~\ref{sec:meth} we describe
chemical and cooling/heating processes included in the radiative
transfer code. In \S~\ref{sec:res} we discuss some representative
simulations and find analytical formulae that fit the simulation
results. In \S~\ref{sec:neg-pos} we use this analytical relationship
to constraint the parameter space where negative or positive feedback
occurs. In \S~\ref{sec:back} we discuss the opacity of the IGM to the
photo-dissociating background. We summarize our results in
\S~\ref{sec:sum}.

Throughout this paper, when we talk about ``small halos'', we refer to
objects with virial temperatures $T_{vir} \la 10^4$ K that rely on the
presence of molecular hydrogen in order to collapse. Conversely,
``large halos'' are objects with $T_{vir} \ga 10^4$ K that can cool
down and collapse even if H$_2$ is completely depleted.  The
cosmological model we adopt is a flat $\Lambda$CDM model with density
parameters $(\Omega_\Lambda, \Omega_0, \Omega_b)=(0.7, 0.3, 0.04)$,
and Hubble constant $H_0=70$ km s$^{-1}$ Mpc\m.  Thus, the baryon
density is $\Omega_bh^2=0.0196$, compatible with $D/H$ inferences from
Big-Bang nucleosynthesis \citep{Burles:98}.

\section{Radiative Transfer in the Primordial
  Universe\label{sec:meth}}

We use 1D radiative transfer simulations to investigate the evolution
of H$_2$ photo-dissociation regions in the primordial universe.  We
are interested in studying the evolution of the abundances and
ionization states of hydrogen, helium, and molecular hydrogen around a
single ionizing/dissociating source that turns on at redshift $z_i$.
Hubble expansion and Compton heating/cooling are also included.

\subsection{1D Non-equilibrium Radiative Transfer}

The radiative transfer equation in an expanding universe in comoving
coordinates has the following form:
\begin {equation}
{1 \over c}{\partial J_{\nu}(R) \over \partial t}+
{1 \over a}{\partial J_{\nu}(R) \over \partial R}-
{H \over c}\left({\partial J_{\nu}(R) \over \partial \ln \nu}-3
  J_{\nu}(R) \right)=\epsilon_\nu - \kappa_\nu J_{\nu}(R),\label{eq:rt}
\end {equation}
where $J_{\nu}(R)$ [erg s\m cm\mm Hz\m sr\m] is the specific intensity
and $a=(1+z_{em})/(1+z_{abs})$,
with $z_{em}$ and $z_{abs}$ the emission and absorption redshifts, and
$\epsilon_\nu$ and $\kappa_\nu$ the emissivity and absorption
coefficients. 
If the scale of the problem is smaller than the horizon size, $c/H$,
and $|\partial J_{\nu}/ \partial \ln \nu | \sim J_{\nu}$ (true
in the case of continuum radiation), we can neglect the cosmological
terms. In \S~\ref{ssec:LW} we show that we can also use the classical
radiative transfer equation for lines if we account for the
cosmological redshift of the radiation (or line cross sections).
Finally, if the emission and absorption coefficients do not change on
time scales shorter than a light-crossing time, ($L/c$), we can solve
at each time step the stationary classical radiative transfer
equation. For the case of a single point source, the solution is
the attenuation equation
\begin {equation}
J_{\nu}(R)={J_{\nu}(0) \over R^2}\exp[-\tau_{\nu}(R)],
\end {equation}
where
\begin {equation}
\tau_{\nu}(R)=\sum_i \int_0^R n_i(r) \sigma_i(\nu)~dr,
\end {equation}
with $n_i$ the number density and $\sigma_i$ the absorption cross
section of species $i$.

\subsection{Chemistry and Initial Conditions}

We solve the time dependent equations for the photo-chemical
formation/destruction of 8 chemical species (H, H$^+$, H$^-$, H$_2$,
H$_2^+$, He, He$^+$, He$^{++}$), including the 37 main
processes relevant to determine their abundances \citep{ShapiroKang:87}.
We use ionization cross sections from
\cite{Hui:97} and photo-dissociation cross sections from \cite{Abel:97}.
The energy equation is
\begin{equation}
{d\epsilon_g \over dt}=\Gamma-\Lambda ,
\end{equation}
where $\epsilon_g=(3kT/2) n_H(1+x(He)+x_e)$, with $n_H$ the hydrogen number
density, and $x(He)$ and $x_e$ the helium and electron fractions
respectively. Note that $n_H$ is a function of time because of the
Hubble expansion. Also, $x_e$ is time dependent.
The cooling function $\Lambda$ includes H and He line and continuum
cooling \citep{ShapiroKang:87} and molecular ro-vibrational cooling
excited by collisions with H and H$_2$ \citep{Martin:96,Galli:98}. As
mentioned above, adiabatic cosmic expansion cooling is also included.
The heating term $\Gamma$ includes Compton heating/cooling and
photoionization/dissociation heating.  We solve the system of ODEs for
the abundances and energy equations, using a $4^{\rm th}$-order
Runge-Kutta solver. We switch to a semi-implicit solver
\citep{GnedinG:98} when it is more efficient (\ie, when the
abundances in the grid are close to their equilibrium values).
Convergence analysis showed that a logarithmic grid with 400 cells in
space and 400 cells in frequency (for the continuum flux) coordinates is
sufficient to achieve convergence within a 10\% error. The spectral
range of the radiation is between 0.7 eV and 1 keV.

The primordial helium mass fraction is $Y_P=\rho(He)/\rho_b=0.24$,
where $\rho_b$ is the baryon density, so that
$x(He)=Y_P/4(1-Y_P)=0.0789$. The initial values at $z=z_i$ for the
temperature and species abundances in the IGM are: $T=10$ K, $x_{H_2}=2
\times 10^{-6}$, and $x_e \simeq
x_{H^+}=10^{-5}/(h\Omega_b\Omega_0^{1/2})=6.73 \times 10^{-4}$.  The
initial abundance of the other ions is set to zero.  We explore two
cases: (i) sources embedded in the constant density IGM and, (ii)
sources inside a virialized halo. For case (ii) we use initial
conditions and a density profile to match the numerical simulation of
\cite{Abel:00}.

The density distribution of massive stars in the halo is crucial for
determining \fesc, the escape fraction of ionizing photons
\citep{RicottiS:00}. If all the stars are located at the center of the
DM potential, it is trivial to find \fesc for a given baryonic density
profile, because the calculation is reduced to the case of a single
Str\"omgren sphere,
\begin{equation}
   f_{esc} = 1-{4\pi \alpha_H^{(2)} \over S_0}\int_0^{\infty}
          n_H^2(R)R^2dR \;,
\end{equation}
where $\alpha_{H}^{(2)} = 2.59 \times 10^{-13}$ cm$^3$ s$^{-1}$ is the
case-B recombination rate coefficient at $T=10^4$ K and $n_H$ (cm\mmm) is the
hydrogen number density.  Assuming that the gas profile is given by
solving the hydrostatic equilibrium equation in a DM density profile
\citep{Navarro:97}, and that the Lyc photon luminosity, $S_0$
[photon s\m], is proportional to the baryonic mass of the galaxy, we
find,
\begin{equation}
   f_{esc} = 
\begin{cases}
1-0.55f_g{(1+z)^3 \over \epsilon } & \text{if $(1+z) < 1.22 (\epsilon/f_g)^{1/3}$},\\
0 & \text{if $(1+z) > 1.22 (\epsilon/f_g)^{1/3}$}.
\end{cases}
\label{eq:c1}
\end{equation}
Here, $z$ is the collapse redshift of the halo, $S_0 =(1.14 \times
10^{49}$ s\m)$\epsilon f_g (M_{DM}/10^6$ M$_\odot$), $\epsilon$ is the
star formation efficiency normalized to the Milky Way, and $f_g$ is
the collapsed gas fraction; see \cite{RicottiS:00} for details. We
expect $f_g$ to be small since the DM potential wells of this objects
are too shallow to hold photo-ionized gas.  When \fesc is not zero,
the number of photons absorbed is such that all the gas in the halo is
kept ionized.  Since we assume spherical symmetry, we can only place
the source at the center of the halo.  From the steep rise of \fesc in
equation~(\ref{eq:c1}), it is clear that \fesc is essentially either 1
or 0 for the majority of the redshifts and star formation
efficiencies.  In a more realistic geometry, \fesc turns out to be
small but not zero.  We can make a more realistic calculation if we
place the grid outside the virial radius in the constant-density IGM
and reduce the ionizing flux by a factor \fesc \footnote{For the sake
  of simplicity we simply reduce the ionizing flux by a factor \fesc
  without changing the SED of the source that, in a realistic case,
  should become harder. Indeed, as noted in \S~\ref{sec:res}, in the
  simulations where the source is at the center of a halo the emerging
  ionizing photons are both reduced and harder}.  The escape fraction
from the halo can be estimated with more accurate calculations
\citep{Dove:00,RicottiS:00} or left as a free parameter.

\subsection{Photon Production}

The spectral energy distribution (SED) of the source is assumed to be
either a constant power law, $F_\nu \propto \nu^{-\alpha}$, with
$\alpha=1.8$ for the case of a mini-quasar (QSO), a constant (at
$t=0$) SED from a population of metal-free stars for the Pop~III
source \citep{Tumlinson:00}, or an evolving SED of stars with
metallicity $Z=0.001$ \citep{Leitherer:95} for the Pop~II source.  We
consider the two extreme cases of an instantaneous burst of star
formation or continuous star formation.

In Figure~\ref{fig:sed1} we show the SED of a zero metallicity
population for instantaneous star formation, in which $100$ M$_\odot$
of gas is converted into stars with a Salpeter IMF, a lower mass
cut-off at $M_{low}=1$ M$_\odot$ and upper cut-off at $M_{up}=100$
M$_\odot$.  The solid line shows the non-evolved SED (at $t=0$). We
also show the photo-ionization and photo-dissociation cross sections
for H$^-$ (dotted line), H$_2$ (dashed line), and H$_2^+$ (dot-dashed
line). The H$_2$ photo-dissociation cross section in the Lyman-Werner
bands is the average cross section of the lines between 11.2 and 13.6
eV according to \cite{Abel:97}. In Figure~\ref{fig:sed2} we show the
SED of the $Z=0.001$ metallicity population with the same parameters
as for the Pop~III in Figure~\ref{fig:sed1}.
 
We have explored the effects of changing the IMF and mass cut-off. The main
difference is that the Lyc luminosity $S_0$ can change by as much as an
order of magnitude for a steeper IMF ($\alpha=3.3$) or for a Salpeter
IMF ($\alpha=2.35$) with $M_{up}=30$ M$_\odot$. The Lyc luminosity at
$t=0$ generated by converting instantaneously $10^6$ M$_\odot$ of
baryons into stars is $S_0 \simeq 10^{53}$ [photon s\m] for the
Pop~III SED and $S_0 \simeq 6 \times 10^{52}$ [photon s\m] for the
Pop~II SED. The specific flux from the object is $F_{\nu}=L_{\nu}/4\pi
R^2$ [erg s\m cm\mm Hz\m] where $L_\nu$ is the specific luminosity.
In a previous conference proceedings \citep{RicottiGS:00} we
erroneously underestimated the specific intensity by a factor $4\pi$
for the Pop~II and Pop~III SEDs. Therefore the stated photon
luminosities in those figures should be a factor of $4\pi$ bigger.

\def\capfa{%
  SED of a zero metallicity population for
  instantaneous star formation, in which $100$ M$_\odot$ of gas is
  converted into stars with a Salpeter IMF, a lower mass cut-off at
  $M_{low}=1$ M$_\odot$ and upper cut-off at $M_{up}=100$ M$_\odot$.
  The solid line shows the non-evolved SED (at $t=0$). We also show
  the photo-ionization and photo-dissociation cross sections for H$^-$
  (dotted line),
  H$_2$ (dashed line), and H$_2^+$ (dot-dashed line). The
  H$_2$ photo-dissociation cross section in the Lyman-Werner bands is
  the average cross section of the lines between 11.2 and 13.6 eV
  according to \cite{Abel:97}.}
\placefig{
\begin{figure*}[thp]
\plotone{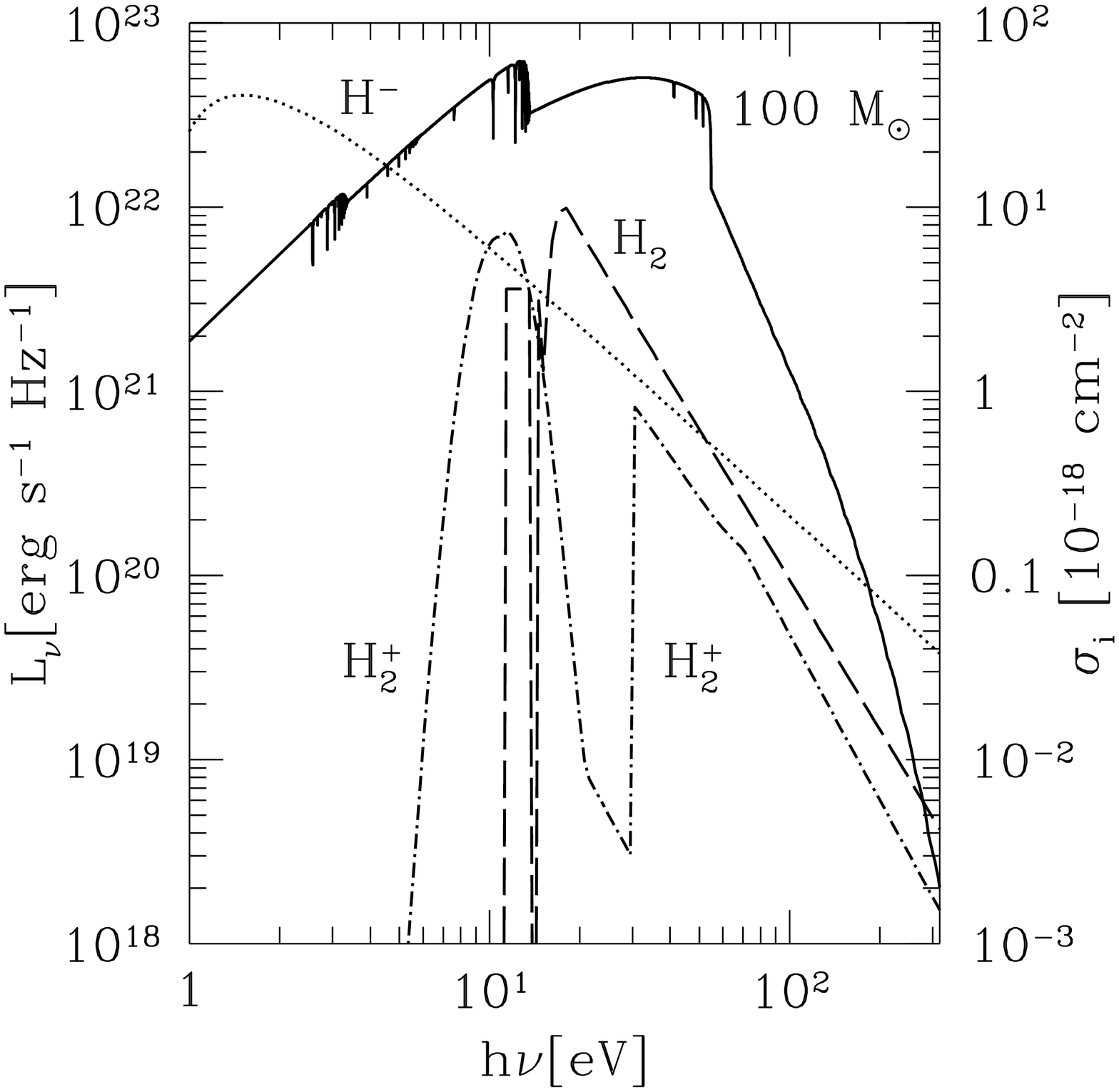}
\caption{\label{fig:sed1}\capfa}
\end{figure*}
} 

\def\capfb{SED of the $Z=0.001$ metallicity population with the
  same parameters as for the Pop~III in Figure~\ref{fig:sed1}. The
  lines show the SED at $t=1, 5, 10, 100$ Myr after the burst
  \citep{Leitherer:95}.}
\placefig{
\begin{figure*}[thp]
\plotone{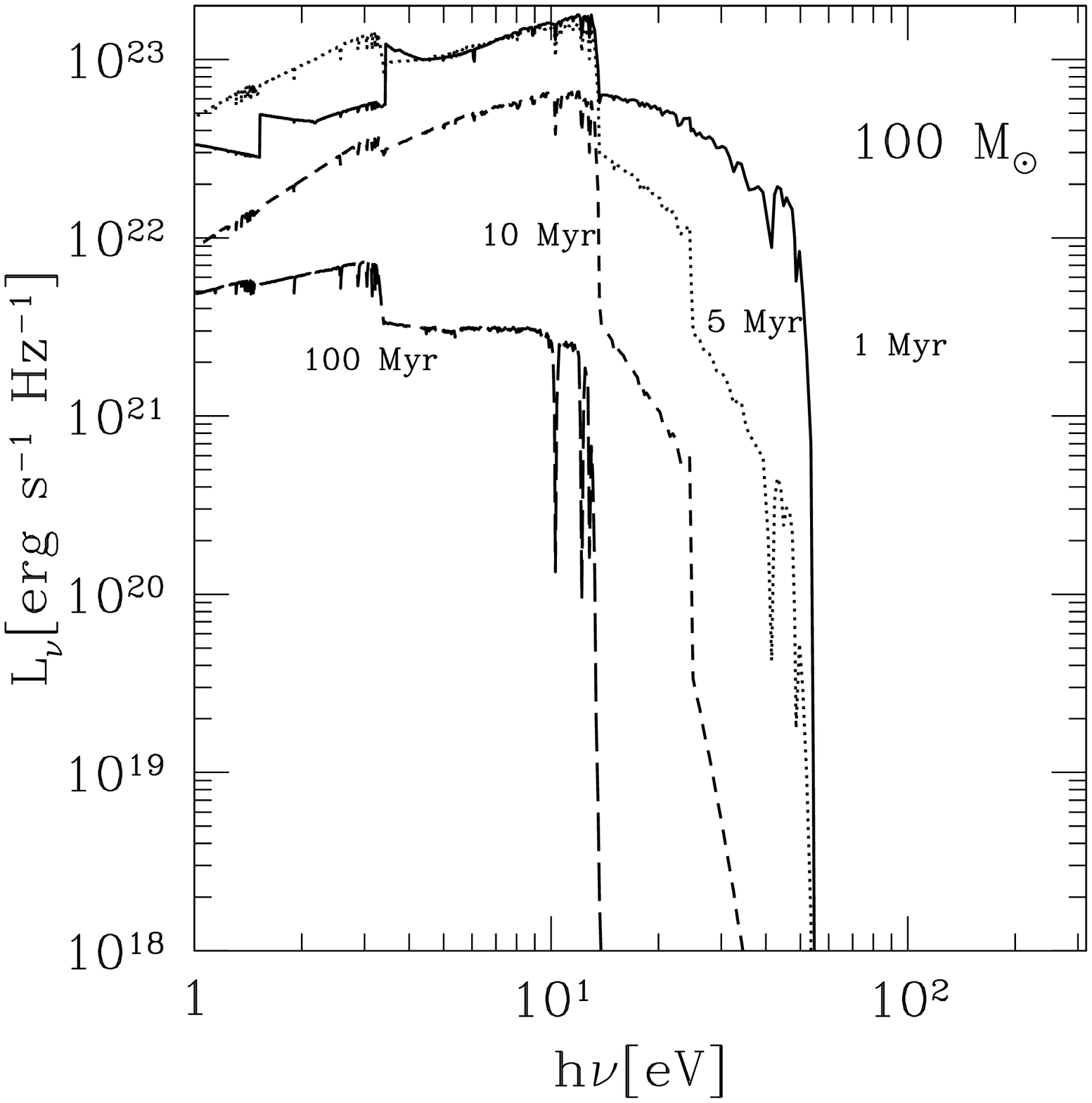}
\caption{\label{fig:sed2}\capfb}
\end{figure*}
}

\subsection{Radiative Transfer for the Lyman-Werner Bands\label{ssec:LW}}

In Figure~\ref{fig:1} we show a representative simulation after $t\sim
16$ Myr from a burst of a Pop~III objects turning on at $z_i=19$ with
total Lyc photon luminosity $S_0=1.2 \times 10^{49}$ s$^{-1}$.  We
find a shell of H$_2$ formation just in front of the H~II region that
we call a {\em positive feedback region} (PFR), where, in some cases,
the abundance of H$_2$ is much higher that the relic abundance of
$x_{H_2}=2 \times 10^{-6}$. Figure~\ref{fig:1b} shows the gas
temperature as a function of the comoving distance from the source at,
$t\sim 0.05,~1$, and 22 Myr for the same object shown in
Figure~\ref{fig:1}. Even if the temperature is $T \sim 10^4$ K in the
PFR, collisional dissociation of H$_2$ is not very effective at the
typical densities of the IGM or in the outer part of galaxy halos.

\def\capfc{Abundances as a function of the comoving distance from
  the source at $t\sim 100$ Myr for a Pop~III object turning on at
  $z_i=30$ with Lyc photon luminosity $S_0=1.2 \times 10^{49}$
  s$^{-1}$.  The H and He~I ionization fronts are at $R \simeq 5$ kpc
  and just ahead of the fronts is the PDF region with an H$_2$
  abundance $x_{H_2} \simeq 5 \times 10^{-4}$. The thick (thin) line shows
  $x_{H_2}$ including (excluding) line radiative transfer in H$_2$
  Lyman-Werner bands.}
\placefig{
\begin{figure*}[htp]
\plotone{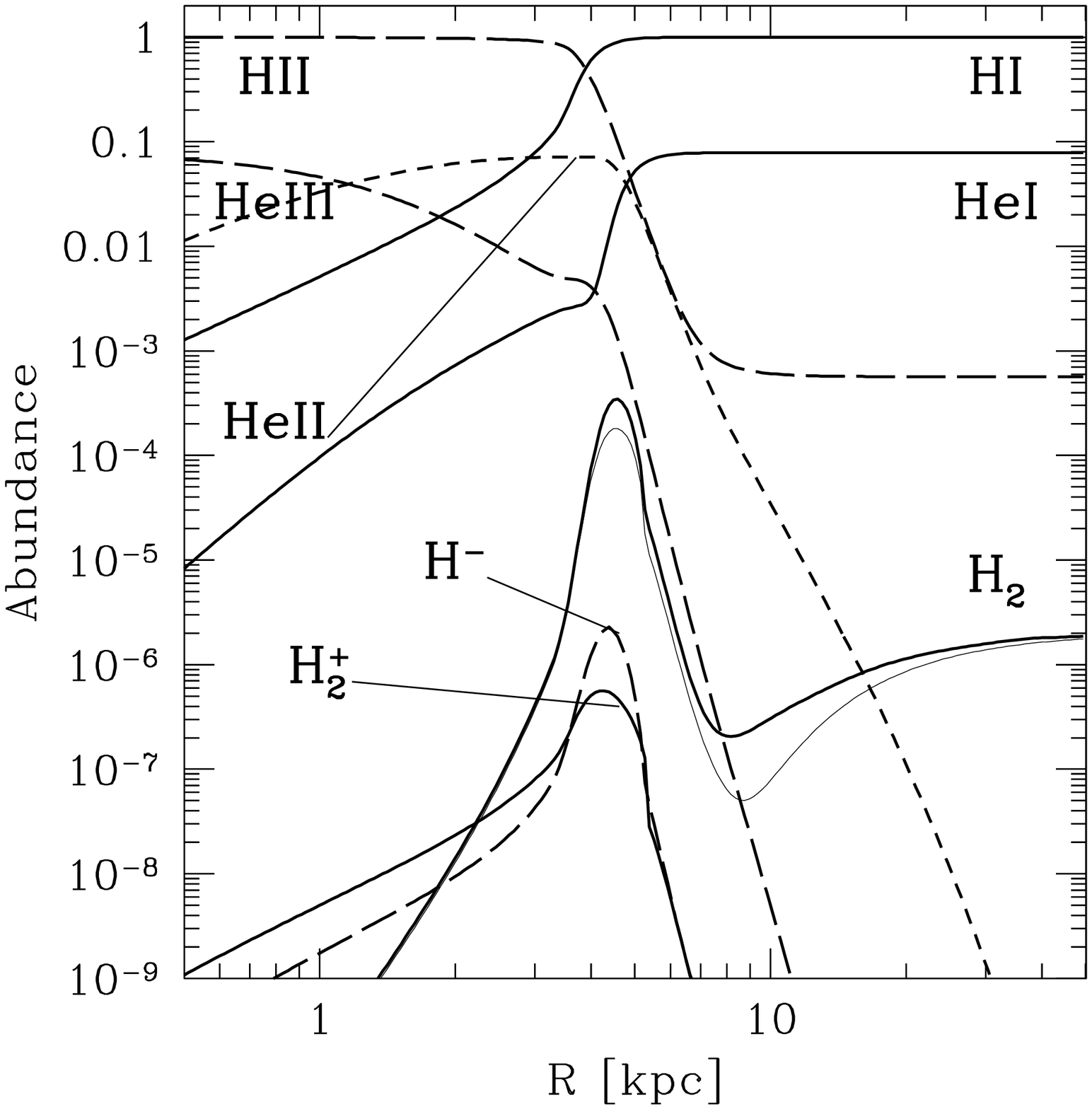}
\caption{\label{fig:1}\capfc}
\end{figure*}
}
\def\capfd{Gas temperature as a function of the comoving distance
  from the source at times $t\sim 0.05,~1$, and 22 Myr after a Pop~III
  object turns on at $z_i=19$ with Lyc photon luminosity $S_0=1.2
  \times 10^{49}$ s$^{-1}$.}
\placefig{
\begin{figure*}[htp]
\plotone{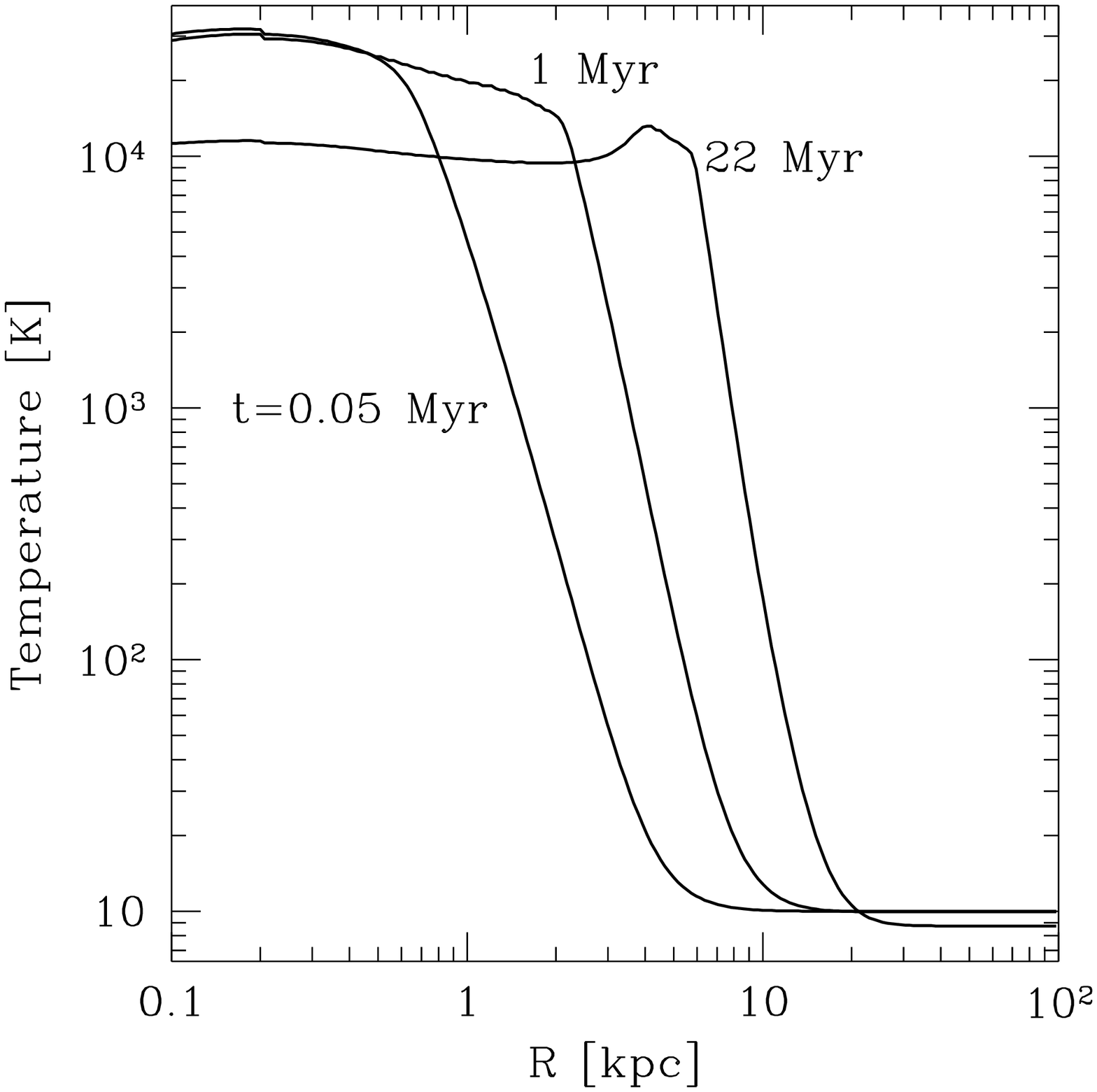}
\caption{\label{fig:1b}\capfd}
\end{figure*}
}

\def\capfe{Emerging spectrum in the Lyman-Werner bands at $t=50$
  Myr after a Pop~III object turns on
  at $z_i=30$ with Lyc photon luminosity $S_0 \simeq 10^{49}$
  s$^{-1}$.  Each panel shows the spectrum at progressively higher
  resolution. We only show the Lyman series absorption lines of
  hydrogen that are important in reducing the photo-dissociation rate
  of H$_2$. The molecular absorption lines are produced by the
  optically thick PFR. The solid line shows the region of the spectrum
  in which we calculate line radiative transfer. The dotted line shows
  the SED of the source.}
\placefig{
\begin{figure*}[htp]
\plotone{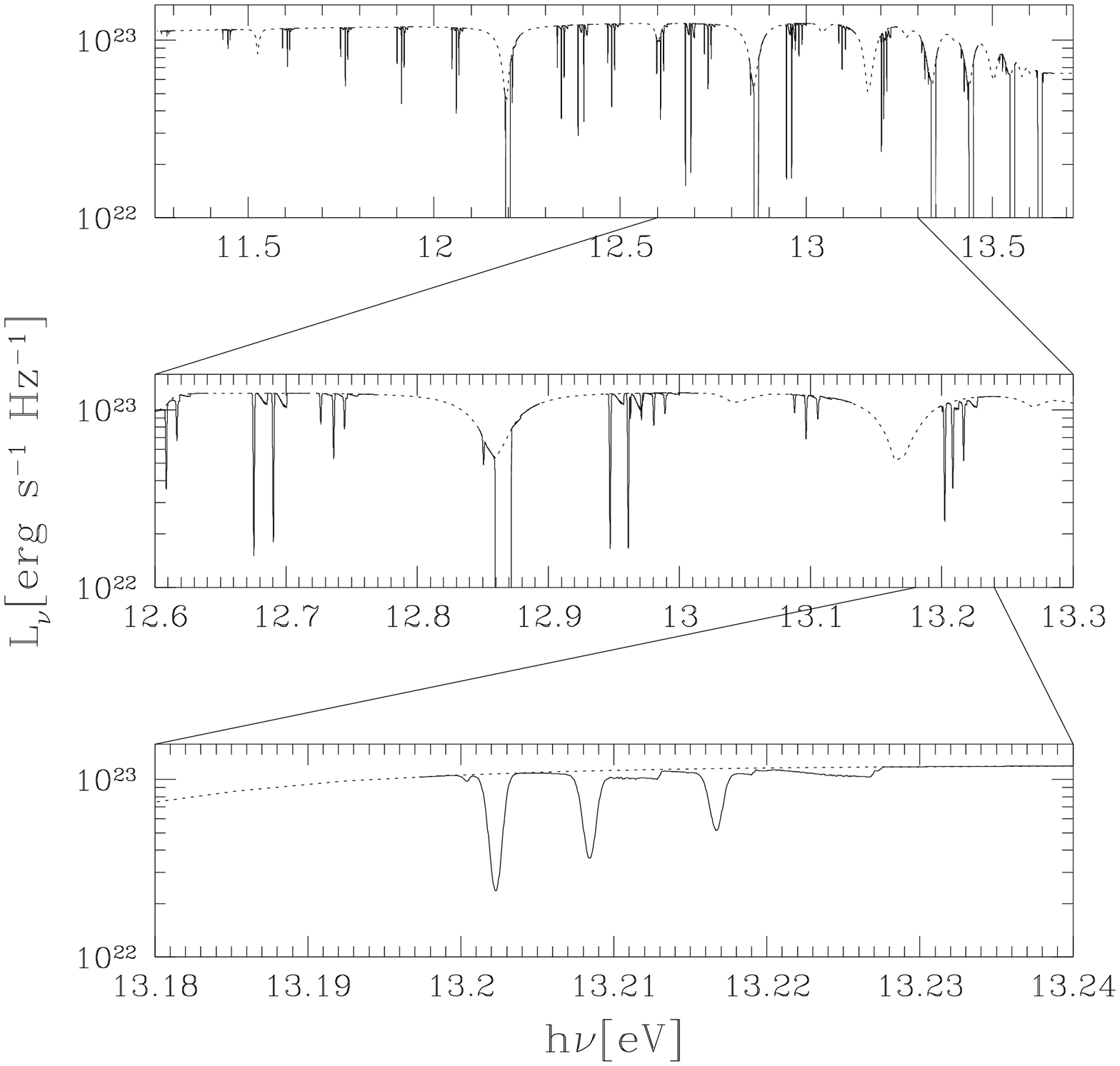}
\caption{\label{fig:LWspec}\capfe}
\end{figure*}
}

In other cases, we find that the column density of H$_2$ formed ahead
of the ionization region is high enough ($N \ga 10^{14}$ cm$^{-2}$)
that some Lyman-Werner bands are optically thick at line center. In
order to properly include this effect, we solve the radiative
transfer, not only for the continuum radiation but also in lines.
Since the H$_2$ and H~I lines are very narrow (because the IGM
temperature is $T\sim 10$ K [$\Delta \nu_{th}/ \nu \sim 10^{-6}]$), we
need extremely high spectral resolution in the frequency range of
$10.2-13.6$ eV. Fortunately, we only need to resolve 76 absorption
lines longward of $912$ \AA\ [R(0), R(1), P(1) Lyman lines and R(0),
Q(1), R(1) Werner lines] arising from the J=0 and J=1 rotational
levels of $v=0$ in the H$_2$ ground electronic state. Thus, we only
need high spectral resolution in the vicinity of the absorption lines.
This limits the number of the frequency bins to just $3 \times 10^4$
instead of $\sim 10^6$ for the uniform sampling.  At densities $n_H
\la 10$ cm\mmm, the population of upper rotational or vibrational
states of H$_2$ is negligible even at $T =10^4$ K. We also include
some H~I Lyman-series resonance lines that are close in frequency to
H$_2$ lines. For these lines we use a Voigt profile, since they are
usually in the damping-wing regime of the curve of growth. For the
Lyman-Werner lines we use Gaussian profiles since the optical depth
does not exceed a few. We redshift the specific flux according to the
distance from the source, so that the maximum redshift of a line
depends on the grid physical size. For our simulation the maximum
frequency redshift is smaller than the distance of two consecutive
Lyman or Werner line ``triplets''.  Therefore we need to consider only
a limited frequency range around each triplet.  We resolve the
frequency profile of each one of the 76 absorption lines in the
Lyman-Werner bands with about 100 grid points if $T \sim 10^4$ K and
with a few points if $T \sim 10$ K. Another complication arises from
the fact that our spatial grid is logarithmic. Therefore, at a large
enough distance from the source, the line can be redshifted several
line widths inside each grid cell.  We will show in \S~\ref{sec:back}
that, for the simple case of constant density, temperature, and
molecular abundance, the optical depth can be calculated analytically.
The line profile, in this case, is similar to a top-hat function with
$\tau=x_{H_2}n_H(z) \sigma f_{osc,i} \lambda_i/H(z)$ if $\nu_i-\Delta
\nu \le \nu \le \nu_i$ and zero otherwise, where $\Delta\nu=H(z)\Delta
x /\lambda_i$ (see Figure~\ref{fig:oneline}). Here, $H(z)$ is the
Hubble constant, $\sigma=(\pi e^2/m_e c)$ is the classical cross
section, $f_{osc,i}$ is the oscillator strength, and $\lambda_i$ is
the wavelength of the $i^{th}$ line.  In Figure~\ref{fig:LWspec} we
show an example of the Lyman-Werner bands spectrum emerging from the
PFR after $t=50$ Myr from the source turn-on. The source is a Pop~III
object with Lyc photon luminosity $S_0 \simeq 10^{49}$ s$^{-1}$
turning on at $z=30$.  Each panel shows the spectrum at progressively
higher resolution. 

\section{Results\label{sec:res}}

\subsection{Simulation Results \label{ssec:sim}}

Preceding the ionization front, a thick shell of several kpc of
molecular hydrogen (with $x_{H_2}$ up to $10^{-4}$) forms
because of the enhanced electron fraction in the transition region
from the H~II region to the neutral IGM. This shell can be optically
thick in some Lyman-Werner H$_2$ lines depending on the redshift, source
luminosity, and escape fraction. This has two main consequences: (1)
the photo-dissociation fronts around the sources slow down and their
final radii are smaller than in the optically thin case; (2) optically
thick PFRs could reduce the intensity of the cosmological background in the
Lyman-Werner bands.

In Figures~\ref{fig:2}-\ref{fig:3} we show the isocontours of
$\log(x_{H_2})$ as a function of time and comoving
distance from the source.  The thick lines show the analytical fits
for the position of the photo-dissociation, formation and ionization
fronts as a function of time.  Figure~\ref{fig:2}~(top) shows an
example where positive feedback is unimportant. The emitting object
has a Pop~II SED, $S_0 \simeq 10^{50}$ s\m, and a turn-on redshift
$z_i=30$.  The source is at the center of a baryonic halo with density
profile $n \propto R^{-2}$. Figure~\ref{fig:2}~(bottom), instead,
shows an example where positive feedback is important. The source has
Pop~III SED, $S_0 \simeq 10^{50}$ s\m, and a turn-on redshift
$z_i=20$. The photo-dissociation front moves so slowly that the PFR
shell gets close to it, producing an enhancement of H$_2$ abundance
instead of a depletion.  In Figure~\ref{fig:2a}~(top) we show a QSO
SED, $S_0 \simeq 10^{51}$ s\m, and a turn-on redshift $z_i=20$. The
mini-quasar produces fronts similar to the Pop~III object. In
Figure~\ref{fig:2a}~(bottom) the source is a Pop~II object with
instantaneous burst of star formation. After $t \sim 10^7$ yr, when OB stars
start to explode as supernovae, the H~II region recombines, triggering
the formation of H$_2$.  Finally,
Figure~\ref{fig:3}~(bottom) shows the effect of including radiative
transfer in the Lyman-Werner lines; the photo-dissociation front slows
down considerably with respect to the optically thin case (top).

The UVB background affects these results in a simple way. The H$_2$
abundance is $x_{H_2} \propto 1/(F_s+F_{LW})$ where $F_s$ and $F_{LW}$ are
the fluxes in the Lyman-Werner band from the source and from the
background respectively.  In the optically thin case $F_s \sim 8.4
\times 10^{-22} R_{\rm kpc}^{-2}(S_0/10^{49}~{\rm s}^{-1})$ [erg s\m
cm\mm Hz\m]. Therefore, the background flux equals the source flux at a
distance from the source $R_{crit} \sim (8.4 \times
10^{-22}/F_{LW})^{1/2}(S_0/10^{49}~{\rm s}^{-1})^{1/2}$ kpc. At a distance $R
= 3 R_{crit}$ the background flux will
reduce our calculations of $x_{H_2}$ by a factor of 10.

\def\capff{Contours of log$(x_{H_2})$ as a
  function of time and comoving distance from the source. The thick
  lines show the analytical fits of the photo-dissociation front (D),
  ionization front (I) and formation region (F). (Top) example of
  negative feedback: Pop~II SED, $S_0 \simeq 10^{50}$ s\m, turn-on
  redshift $z_i=29$. The source is at the center of baryonic halo with
  density profile $n \propto R^{-2}$. (Bottom) example of positive
  feedback: Pop~III SED, $S_0 \simeq 10^{50}$ s\m, turn-on redshift
  $z_i=19$.}
\placefig{
\begin{figure*}[hpt]
\epsscale{0.7}
\plotone{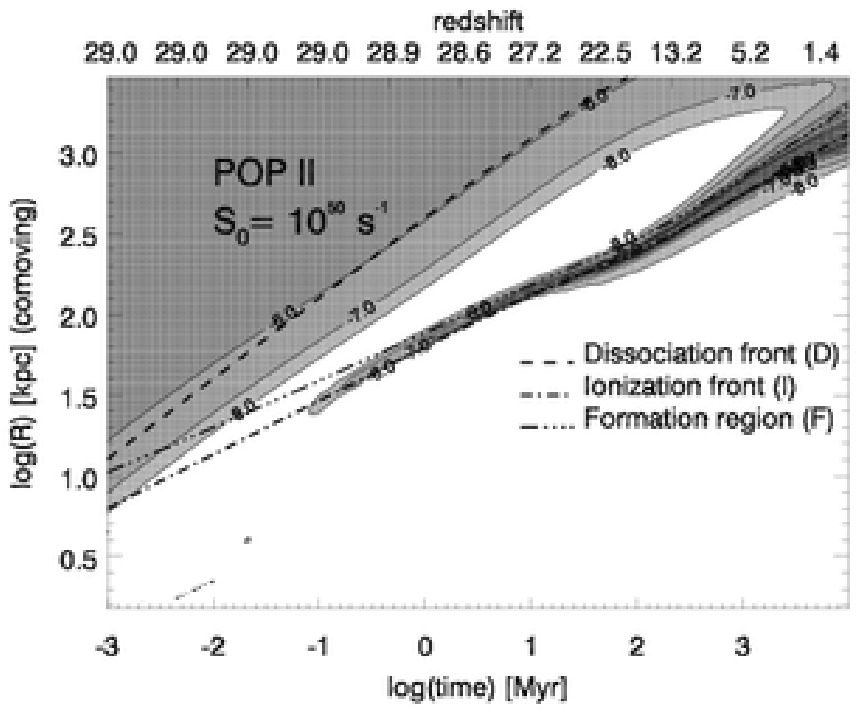}
\plotone{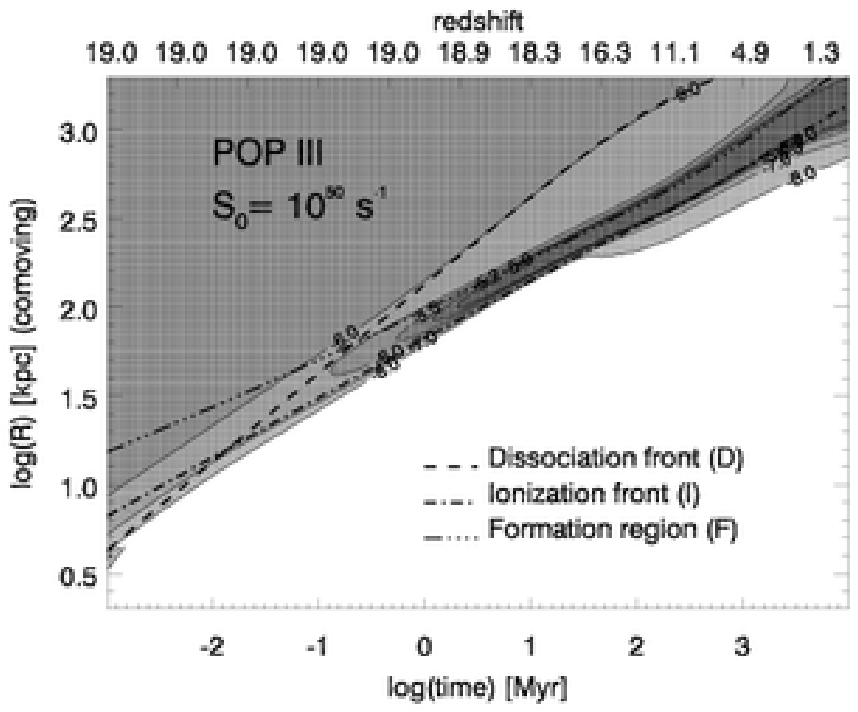}
\caption{\label{fig:2}\capff}
\end{figure*}
}

\def\capfg{Same as Figure~\ref{fig:2}. (Top) QSO SED, $S_0 \simeq
  10^{51}$ s\m, turn-on redshift $z_i=19$. (Bottom) fossil H~II Region:
  Pop~II SED (instantaneous star formation law),  $S_0 \simeq
  10^{49}$ s\m, turn-on redshift $z_i=14$.}
\placefig{
\begin{figure*}[hpt]
\epsscale{0.7}
\plotone{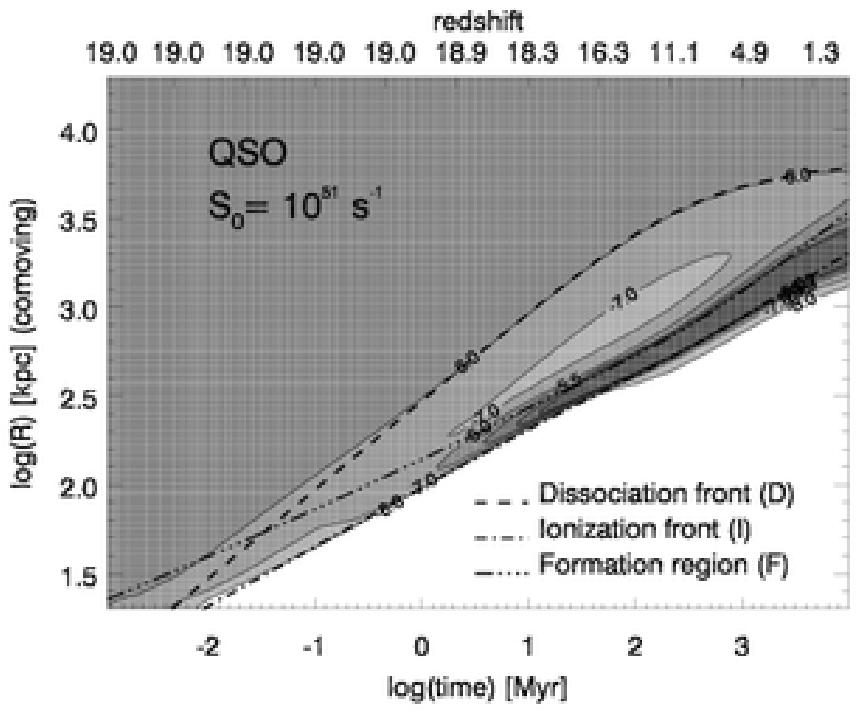}
\plotone{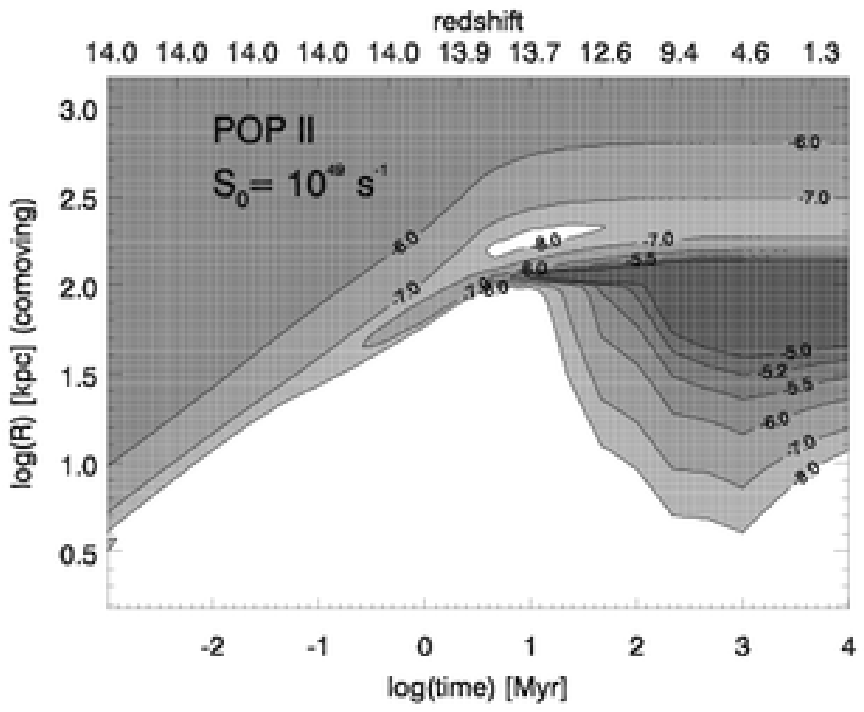}
\caption{\label{fig:2a}\capfg}
\end{figure*}
}
\def\capfh{Same as Figure~\ref{fig:2}. (Top) Pop~III SED, $S_0 \simeq
  10^{49}$ s\m, turn-on redshift $z_i=30$. (Bottom) same but including
  line radiative transfer in H$_2$ Lyman-Werner bands.}
\placefig{
\begin{figure*}[hpt]
\epsscale{0.7}
\plotone{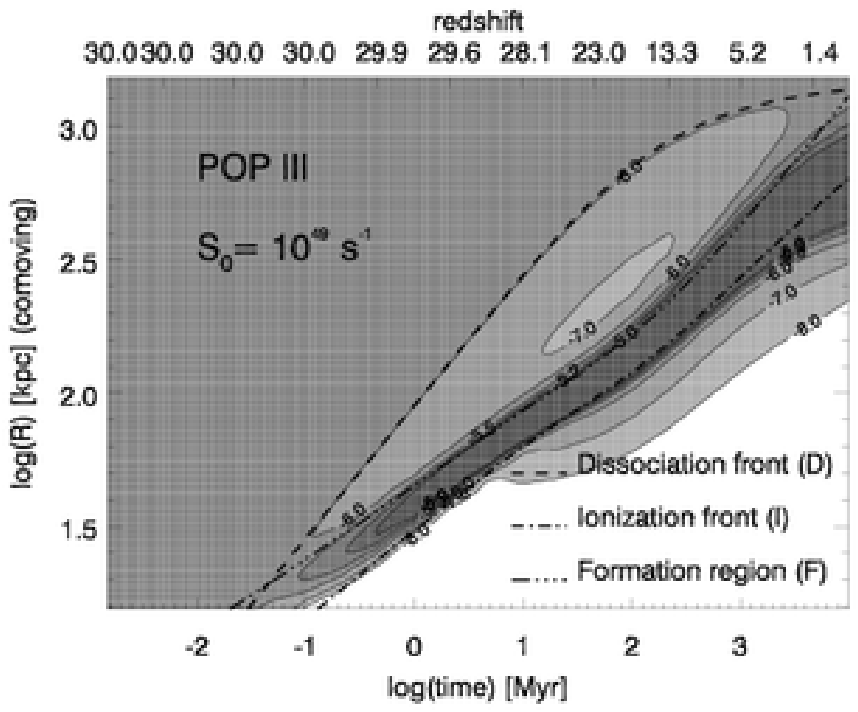}
\plotone{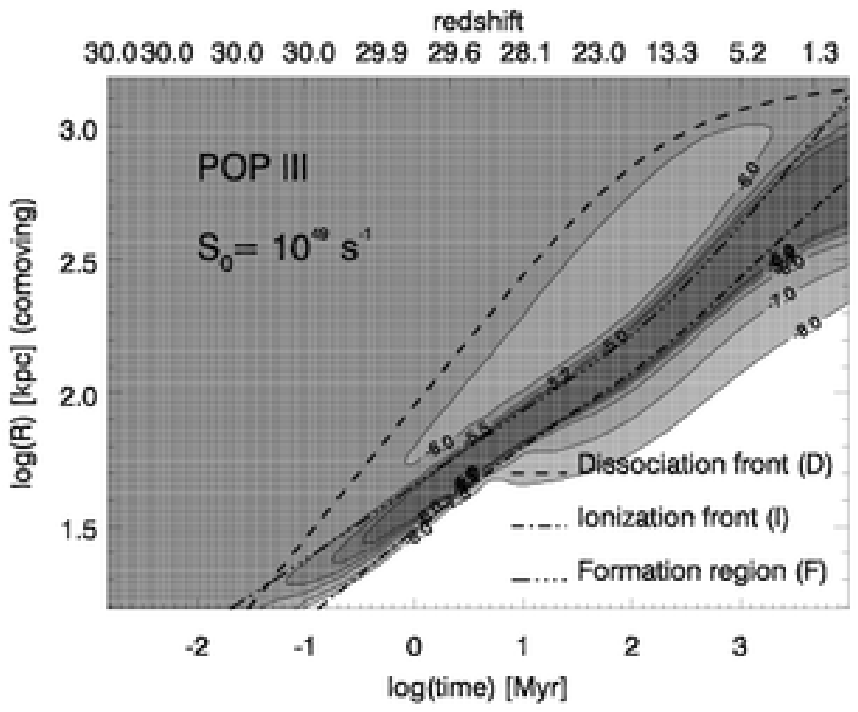}
\caption{\label{fig:3}\capfh}
\end{figure*}
}

\subsection{Analytical Fits \label{ssec:anal}}

Let us introduce some relevant time scales. The collisional
recombination time scales for neutral hydrogen and protons are
$t_{HI}=t_{rec}(1-x_e)/x_e^2$ and $t_{HII}=t_{rec}/x_e$ respectively,
where $x_e$ is the electron fraction number density.  Here we define
$t_{rec}=1/(\alpha_H^{(2)} n_H)$ where $\alpha_{H}^{(2)} = 2.59 \times
10^{-13}$ cm$^3$ s$^{-1}$ is the case-B recombination coefficient at
$T=10^4$ K and $n_H(z_i)=(1.7 \times 10^{-7}
~\text{cm}^{-3})(1+z_i)^3$ is the hydrogen number density for
$\Omega_b h^2 = 0.019$.  The age of the universe $t_H$ and $t_{rec}$
at the redshift $z_i$ when the source turns on are:
\begin{eqnarray}
t_H &\approx& {2 \over 3}{1 \over H_0 \sqrt{\Omega_0}}(1+z_i)^{-3/2}= 538
\left({1+z_i \over 10}\right)^{-3/2}~\text{Myr},\\
t_{rec} &\approx& {1 \over \alpha_H^{(2)} n_H(z_i)}= 720
\left({1+z_i \over 10}\right)^{-3}~\text{Myr}.
\end{eqnarray}
If the density of the IGM is constant and uniform, the
radius $R_I$ of the ionization front is,
\begin{equation}
{R_I \over R_S}=\left[1-\exp\left(-{t \over t_{rec}}\right)\right]^{1 \over 3},
\end{equation}
where $R_S=(3S_0 t_{rec}/4\pi n_H(z_i)]^{1/3}$ is the Str\"omgren
radius and $t$ is the time the source is on.  The typical time for the
H~II region to reach the Str\"omgren radius in a non-expanding IGM
equals the H$^+$ recombination time.  A source turning on at redshifts
$z_i \la 18$, for which $\lambda=t_H/t_{rec} \la 2$, produces an H~II
region that never reaches the Str\"omgren radius \citep{Donahue:87,
  Shapiro:87}. In comoving coordinates, it expands forever because of
cosmic dilution.  This is true only if the source continues to shine
with constant photon luminosity $S_0$. We therefore remind the reader
of another typical time scale, namely the time $t_{burst} \simeq
5-20$~Myr during which an instantaneous burst of star formation
produces a substantial amount of ionizing photons.

Based on the simulations, we find analytical relationships that are
good fits to the comoving front radii (radius of the ionization
sphere, $R_I$, radius of the photo-dissociation sphere, $R_D$, and
outer radius of the PFR, $R_F$), the peak H$_2$ abundance $x_{H_2,F}$,
comoving thickness, $\Delta R_F$ and column density, $N_F$, of the PFR
as a function of time. We also estimate the time scale, $t_F$, on
which a region in the IGM lies inside the PFR. These relations, without
taking into account line radiative transfer, are:
\begin{eqnarray}
R_D &=& (16 ~{\rm kpc})\left({1+z_i \over 10}\right)[\beta S_{0,49} f_2(t)]^{1 \over 2},\label{eq:radius}\\
R_I &=& (21 ~{\rm kpc})[S_{0,49} \langle f_{esc}\rangle f_1(t)]^{1 \over 3},\\
\Delta R_I &=& (5 \times 10^{-5}\gamma~{\rm kpc}){(1+z)^{2.3} \over n_H(z)}(S_{0,49}\langle
f_{esc}\rangle t)^{0.18}\\
&=&(3.2 \gamma~{\rm Mpc})\left({1+z_i \over 10}\right)^{0.7}{(S_{0,49}\langle
f_{esc}\rangle t)^{0.18} \over t_c^{0.466}},\\
\Delta R_F &=& {\Delta R_I \over 3}, \label{eq:drf}
\end{eqnarray}
where $S_{0,49}$ is the ionizing photon luminosity in units of
$10^{49}$ [photon s\m], $\beta$ and $\gamma$ are fitting parameters
(Table~\ref{tab:1}) that depend on the SED of the object
($\beta=\gamma=1$ for a Pop~III SED), and $t$ is the time the source
was on in Myr. $\Delta R_I$ is the comoving thickness of the
ionization front. The functions $f_1$ and $f_2$ are given by
\begin{eqnarray}
f_1(t)&=&{(8.95 \times 10^5)\lambda \over (1+z_i)^3} \exp{\left({\lambda \over
      t_c}\right)}(t_c E_2(\lambda/t_c)-
      E_2(\lambda))~\text{Myr},\\
f_2(t)&=&3t_H[1-t_c^{-1/3}]~\text{Myr},
\end{eqnarray}
with $t_c=1+t/t_H$, $\lambda=t_H/t_{rec}$ and $E_n(x)$ is the
exponential integral of order $n$ \citep{Donahue:87, Shapiro:87}. Note
that $f_1(t)\sim t$ if $t \ll t_H$ and $f_2(t)\sim t$ if $\lambda \la
2$ (\ie $z_i \la 18$). Table~\ref{tab:1} shows the normalization
parameters $\beta, \gamma$ and the ionizing photon luminosity
$S_{0,49}$
produced by an instantaneous burst of $\sim 250$ M$_\odot$ of gas into
stars for the SEDs from Pop~III, Pop~II (continuous star formation)
and mini-quasars.  
\def\tabone{
\begin{deluxetable}{lcccc}
\tablecaption{Fit parameters for the analytical formulae in \S~\ref{ssec:anal}\label{tab:1}}
\tablewidth{0pt}
\tablehead{
\colhead{Object} & \colhead{$S_{0,49}$} & \colhead{$\beta$} & \colhead{$\gamma$}\\
\colhead{} & \colhead{(250 M$_\odot$ burst)} & \colhead{} & \colhead{}}
 \startdata
 Pop~II  & 2.8 & 2.4 & 0.67 \\
 Pop~III & 3.2 & 1.0 & 1.0 \\
 Quasar  & --  & 0.8 & 1.17 \\
 \enddata
\end{deluxetable}
}
\placefig{\tabone}

Here the photo-dissociation front is defined as the locus where the
molecular abundance is $x_{H_2} = 10^{-6}$, half the relic H$_2$
fraction. In the optically thin regime, that is a good approximation
in this case, the profile of the H$_2$ abundance inside the
dissociation sphere is exponential: $x_{H_2} = 2 \times 10^{-6}
\exp[-\ln(2)R_D/r]$. The photo-dissociation front slows down at $t \ga
t_H$, after the initial $t^{1/2}$ expansion law. At late times the
dissociation front approaches the maximum radius $R_D=(340
\alpha)~{\rm kpc} [(1+z_i)/ 10]^{1/4}(\beta S_{0,49})^{1/2}$, where
$\alpha=[1-(1+z_i)^{-1/2}]^{1/2}$.  The relation
$(1+z)=(1+z_i)t_c^{-2/3}$ is used to relate time and redshift.
The maximum thickness of the ionization front $\Delta R_I=(7 \gamma
~\text{Mpc})[(1+z_i)/10]^{0.43}(S_{0,49}\langle
f_{esc}\rangle)^{0.18}$ is reached at $t=0.6 t_H$.

When the ionization front is inside the halo with a density profile $n
\propto R^{-2}$, where $R$ is the radius, we find $\log R_I \propto
(\log t)^2$. The emerging spectrum in this case is harder. The
photo-dissociation front, if we do not take into account
self-shielding in H$_2$ lines, is
independent of the chosen density profile of the gas. The width of the
PFR inside the halo is smaller by a factor $n^{-1}$, but its column
density increases as $n$. 

We have made some numerical ``experiments'' to test which species have
abundances equal to their equilibrium values. Only H$^-$ can be safely
considered in equilibrium. Nevertheless, for H$_2$ the equilibrium
abundance is not a good approximation, it is an helpful
estimate since can be derived analytically \citep{Donahue:91, Abel:97}
and will give us the functional form for the fitting formula.
If we consider only the formation process through H$^{-}$, with rate
$k_f(T) x_{H^-}$ and photo-dissociation by photons with $h\nu> 0.755$
eV, the equilibrium H$_2$ abundance in the PFR is $x_{{H_2},F} \propto
n_H(z) k_f(T)x_{H^-}F_s^{-1} \propto (1+z)R_I^2/(\beta S_{0,49})$, where
$F_s \propto \beta S_{0,49} (1+z)^2 R_I^{-2}$ is the dissociating flux.
In all the simulations, after the PFR shell starts to build up, the
abundance of H$^{-}$ reaches an equilibrium value $x_{H^-}\simeq 5
\times 10^{-7}$.  The best fit differs slightly from the expression
above because of a redshift dependence of the formation rates:
\begin{eqnarray}
x_{{H_2},F} &=& (1.74 \times10^{-9}){(1+z)^{1.3} \over  \beta
 S_{0,49}} R_I^2\\
&=&1.6 \times 10^{-5}\left({1+z_i \over
 10}\right)^{1.3}\beta^{-1}S_{0,49}^{-1/3} \langle
 f_{esc}\rangle^{2/3} {t^{2/3} \over t_c^{0.87}}.
\label{eq:xh2}
\end{eqnarray}
The maximum value of $x_{{H_2},F}$, reached at $t=(10/3)
t_H$, is given by
\begin{equation}
x_{{H_2},F}^{max} = 5.9 \times 10^{-4} \left({1+z_i \over
 10}\right)^{1/3}S_{0,49}^{-1/3} \langle f_{esc}\rangle^{2 \over 3}.
\end{equation}

From equations~(\ref{eq:radius}-\ref{eq:xh2}) it is easy to derive
useful relationships to quantify the importance of positive feedback as a
function of the unknown free parameters. We will use these
relationships in the next section. The ratio
\begin{equation}
{R_D \over R_I}= 0.76 \left({1+z_i \over 10}\right)\langle
f_{esc}\rangle^{-1/3} \beta^{1/2}S_{0,49}^{1/6}\,g(t)
\end{equation}
reaches the maximum value of
\begin{equation}
\left({R_D \over R_I}\right)^{max}= 1.6 \left({1+z_i \over 10}\right)^{3/4}\langle
f_{esc}\rangle^{-1/3} \beta^{1/2}S_{0,49}^{1/6}
\end{equation}
at $t \simeq 0.9 t_H$ where the function
$g(t)=[f_1(t)]^{1/2}[f_2(t)]^{-1/3}$ scales as $g(t) \simeq t^{1/6}$
if $t<t_H$ and $\lambda<1$. The function has a maximum value of
$g=0.75 t_H^{1/6}=2.1 [(1+z_i)/10]^{-1/4}$ at $t=0.9 t_H$.

It is also useful to estimate the H$_2$ column density, $N_F$, of the
PFR, and the time scale, $t_F$, on which a region in the IGM can be
engulfed by the PFR:
\begin{equation}
N_F \simeq {n x_{{H_2},F}\Delta R_F \over 1+z}= (1.6 \times
 10^{13}~\text{cm}^{-2}){\gamma \over \beta}\left({1+z_i \over
 10}\right)^{2.6}{(\langle f_{esc}\rangle t)^{0.85}
 \over S_{0,49}^{0.15}~t_c^{1.7}}
\end{equation}
with the maximum value of
\begin{equation}
N_F^{max}=(9 \times 10^{14}~\text{cm}^{-2}){\gamma \over \beta}\left({1+z_i \over
 10}\right)^{1.3}{\langle f_{esc}\rangle^{0.85} \over
 S_{0,49}^{0.15}}
\end{equation}
at $t \simeq 0.9 t_H$, and
\begin{equation}
t_F = \Delta R_F \left({dR_I \over dt}\right)^{-1}= (152t \gamma)\left({1+z_i \over 10}\right)^{0.7}{(S_{0,49}\langle
f_{esc}\rangle t)^{-0.15} \over t_c^{0.47}}.
\end{equation}

Finally, we derive the ratio $R_D/R_F=R_D/(R_I+\Delta
R_F)=(R_D/R_I)(1+\Delta R_F /R_I)^{-1}$ that gives us an idea of the
fraction of the IGM volume in which the molecular hydrogen is destroyed:
\begin{equation}
{R_D \over R_F} = {R_D \over R_I}\left[1+51 \gamma \left({1+z \over 10}\right)^{0.7}(S_{0,49}\langle
f_{esc}\rangle t)^{-0.15}\right]^{-1}
\end{equation}
with the maximum value of
\begin{equation}
\left({R_D \over R_F}\right)^{max} = {0.15 \beta^{1/2} \over \gamma} \left[\langle f_{esc}\rangle \left({1+z_i \over
        10}\right)\right]^{-0.18}S_{0,49}^{1/3}
\end{equation}
at $t \simeq 0.9 t_H$.

\section{Discussion: Negative or Positive Feedback ? \label{sec:neg-pos}}

\def\capfi{ The contours show $R_D /R_F$ for our standard model and
  cosmological parameters (\S~\ref{sec:int}). The shaded region ($R_D
  /R_F <1$) shows the redshift and halo masses where {\em positive
    feedback} is dominant.  The thick solid line shows the minimum mass
  needed to collapse according to \cite{AbelAnninos:98}. Parameters:
  \fesc= escape fraction, $\epsilon=$ star formation efficiency
  normalized to the Milky-Way, $f_g=$ collapsed gas fraction, $S_0 =
  (1.14 \times 10^{49} {\rm s}^{-1})\epsilon f_g (M_{DM}/10^6$ M$_\odot)$. The
  dot-dashed lines show the collapse redshift of $1\sigma$ and
  $3\sigma$ perturbations according to linear theory. (Top) Pop~III
  SED, $\epsilon=1$, $f_g=1$ and \fesc$=0.2$. (Bottom) Pop~III SED,
  $\epsilon=1$, $f_g=0.1$ and \fesc given by the analytical formula
  derived in \cite{RicottiS:00}.}
\placefig{
\begin{figure*}[htp]
\epsscale{0.65}
\plotone{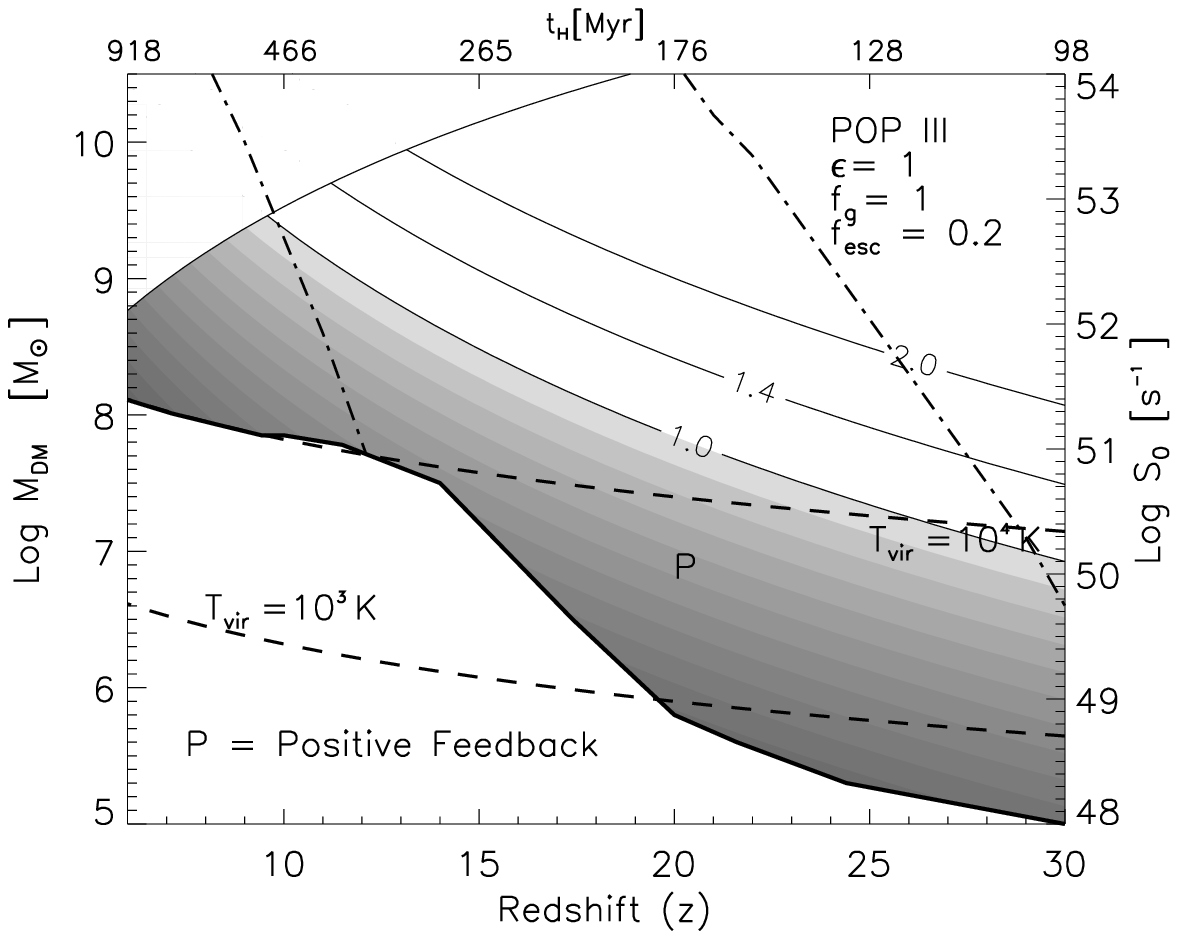}
\plotone{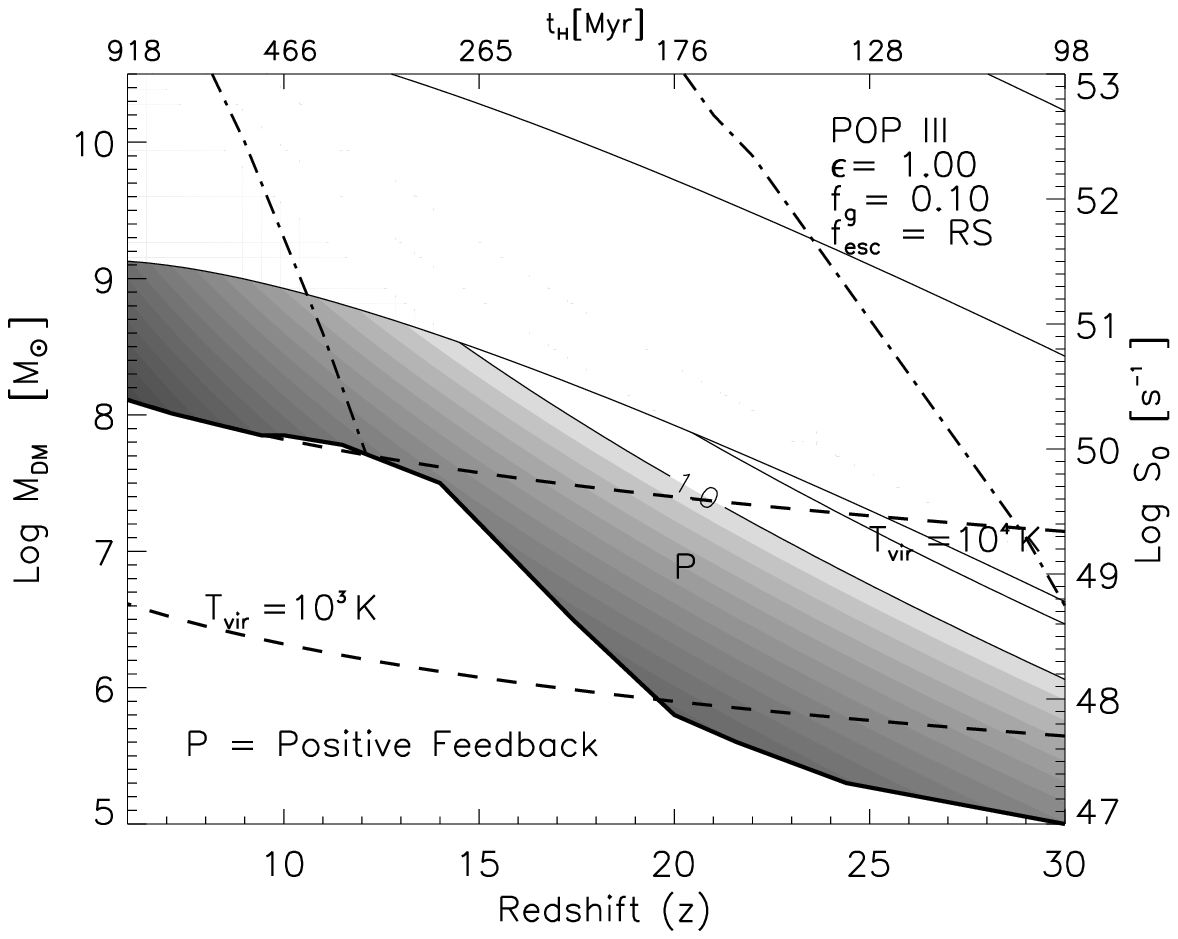}
\caption{\label{fig:4}\capfi}
\end{figure*}
}

In this section, we use the analytical relationships found in
\S~\ref{ssec:anal} to quantify the importance of the positive feedback
as a function of the free parameters of the model. The free parameters
are: the Lyc escape fraction, \fesc, the star formation efficiency
$\epsilon$ normalized to that in the Milky Way, and the collapsed gas
fraction, $f_g$. For a fixed cosmology, the ionizing photon luminosity
is $S_0 = (1.14 \times 10^{49}~\text{s}^{-1}) \epsilon f_g
(M_{DM}/10^6$ M$_\odot)$ [see \cite{RicottiS:00} for details]. We
quantify the importance of the positive feedback by showing in
Figure~\ref{fig:4} isocontours of constant ratio, $R_D/R_F$, of the
photo-dissociation front radius to the formation front radius 10 Myr
after the source turned on. Here the radius of the dissociation front
$R_D$ is defined as the locus where the relic molecular abundance has
dropped to $x_{H_2}=10^{-8}$, therefore it is smaller by a factor
$\ln(200)/\ln(2) \sim 7.6$ with respect to the values given in the previous
paragraph.  Clearly, if $R_D /R_F<1$ the dissociation region does not
exist. Instead, the H$_2$ abundance could increase with respect to the
relic value $x_{H_2} \approx 2 \times 10^{-6}$ (as an example see
Figure 6 [bottom]). Therefore, $R_D /R_F$ gives us an idea of the mean
molecular abundance in the IGM and the opacity of the IGM to the
photo-dissociating background.

We find that the existence of positive feedback depends crucially on
\fesc and the SED of the first objects (Pop~III, Pop~II or AGN).
Figure~\ref{fig:4} shows the parameter space where positive feedback
is possible in the particular case of a Pop~III SED.  The contour
lines show $R_D /R_F$ as a function of redshift and mass of the halo.
The shaded region shows the parameter space where $R_D /R_F<1$, in
which case the PFR fills up the region between the H~II region and the
photo-dissociation front. At small redshifts, the boundary of the
shaded region is produced by the additional constraint that the H$_2$
abundance of the PFR has to be $x_{H_2}>10^{-5}$. The choice of this
value is somewhat arbitrary, but the main purpose is to show the H$_2$
abundance in the PDF as a function of the redshift and halo mass.  The
thick solid line at the bottom is the minimum mass to collapse as a
function of redshift according to the criteria of
\cite{AbelAnninos:98}. The two dashed lines show objects with virial
temperatures of $T_{vir}=10^3$ K and $T_{vir}=10^4$ K; protogalaxies
that lie above $T_{vir}=10^4$ K are not subject to negative/positive
feedback because they can cool by H~I line excitation (Ly$\alpha$).
The minimum mass for collapse is calculated assuming that the object
lies outside of a photo-dissociation region or a PFR and that the
dissociating background is negligible. The amount of H$_2$ formed in a
just-virialized object does not depend on the initial H$_2$ abundance,
but it is sensitive to the electron fraction.  For this reason,
protogalaxies less massive than those found in the
\cite{AbelAnninos:98} calculations can form inside a PFR.  Finally, we
note that the presence of PFRs will increase the opacity of the IGM in
the Lyman-Werner lines, further decreasing the intensity of the
dissociating UV background.  All these effects have to be included in
a 3D cosmological simulation with radiative transfer in order to
understand their global effect on structure formation and the star
formation history.

\section{ IGM optical depth in the Lyman-Werner bands\label{sec:back}}

In this section we estimate the IGM optical depth in the Lyman-Werner
bands that arises from the absorption in 76 narrow H$_2$ lines from
$J=0,1$ in $v=0$ at different redshift.  This calculation is analogous
to the one in \cite{HaimanAR:00}. We repeat it because we disagree
with their technical analysis of H$_2$ line scattering and
fluorescence.  \cite{HaimanAR:00} assumed that only the fraction of
absorptions that decay to the dissociating continuum remove photons
from the UVB (about 11\% of the time).  For the other 89\% of the
absorptions, they assume that the Lyman-Werner photon is re-emitted at
the same frequency, with no net effect. This is not correct. Following
the absorption in the Lyman or Werner bands, the molecule decays to a
variety of ro-vibrational levels in the ground electronic state
because the quadrupole transitions within the electronic excited state
are much slower. The H$_2$ either decays to the $b^3\Sigma^+_u$
(anti-bonding) state which decays to the vibrational continuum
(pre-dissociation) or to a bound vibrational-rotational level in the
electronic ground state ($X^1\Sigma_g^+$). At moderate UV intensities,
the subsequent infrared cascade through the bound levels of the ground
electronic state is entirely determined by the radiative decay rates
\citep{BlackD:76}.  From the cascade probabilities, \cite{Shull:78}
computed that a Lyman-Werner photon has a 14\% probability, on
average, to be re-emitted at the same frequency. Approximately 12\%
of the absorption transitions dissociate and 74\%
fluoresce to excited $(v, J)$ levels of $X^1\Sigma_g^+$.  Roughly
speaking, the probability to decay to one of the other 14 bound
vibrational levels is about 6\%.  Therefore, only 14\% of the time the
Lyman-Werner photon is resonantly scattered following an absorption.
The other 86\% of the time, H$_2$ dissociates or the photon is split into
infrared and less energetic (about 1 eV energy loss) UV photons that
are removed from the Lyman-Werner bands right away or after a few
absorptions.

Therefore, the H$_2$ optical depth of the IGM is given by   
\begin{equation}
\tau_{\nu}(z_{ob},z_{em})={\pi e^2 \over m_e c} \sum_{i=n_1(\nu)}^{n_2(\nu)}
f_{osc,i}(1-f_{i,\nu^{\prime\prime}=0})\int_{z_{ob}}^{z_{em}}
dz^\prime~c {dt \over dz^\prime}~ n_H(z^\prime)x_{H_2}(z^\prime)\phi(\nu
^\prime,\nu_i) 
\label{eq:tau}
\end{equation}
where $z_{ob}$ and $z_{em}$ are the observer and emission redshifts,
$\nu^\prime=\nu(1+z_{em})/(1+z_{ob})$, $n_H(z)=(1.12\times
10^{-5}~\text{cm}^{-3})\Omega_bh^2(1-Y_P)(1+z)^3$ is the neutral H
number density in the IGM, $\phi$ is the line profile, $f_{osc,i}$ is
the oscillator strength, and $f_{i,\nu^{\prime\prime}=0}$ the
probability to decay to the ground vibrational level of the
$X^1\Sigma_g^+$ state for the $i^{th}$ line calculated from
\cite{BlackD:76}.  The maximum redshift interval a UV photon can
travel before it is absorbed by a neutral atom corresponds to the
redshift between two H~I resonances in the higher Lyman series, the ``dark
screen'' approximation of \cite{HaimanAR:00}.  Therefore a photon
is subject to the absorption from a subset of the 76 Lyman-Werner
lines, $n_1(\nu) \le i \le n_2(\nu)$, where $n_1(\nu)$ is the first
Lyman-Werner line with frequency just above $\nu$, and $n_2(\nu)>n_1(\nu)$ is
the Lyman-Werner line with frequency just below the next higher H~I
Lyman line with frequency $>\nu$.
 
It is possible to write an approximate analytical solution of
equation~(\ref{eq:tau}), if we assume Gaussian line profiles,
$\phi_G(\nu,\nu_i)=(1/\sqrt{\pi}\Delta\nu_{i})
\exp[-(\nu-\nu_i)^2/\Delta\nu_{i}^2]$, where $\Delta\nu_{i}=3\times
10^{-7} \nu_i T_{IGM}^{1/2}$ is the Doppler width of the H$_2$ line $i$, or
Lorentzian line profiles
$\phi_L(\nu,\nu_i)=(\Gamma_i/2\pi)/[(\nu-\nu_i)^2+(\Gamma_i/2)^2]$, where
$\Gamma_i=\gamma_i/2\pi$ and $\gamma_i=\sum_l A(i\rightarrow l)$ is the
natural width of the $i^{th}$ H~I line. For a constant $x_{H_2}=2 \times
10^{-6}$ molecular abundance, we have
\begin{equation}
\tau_{\nu}(z_{ob},z_{em})\simeq 3.28 \times 10^{15} \left({1+z_{ob} \over \nu}\right)^{3/2}\sum_{i=n_1(\nu)}^{n_2(\nu)}
f_{osc,i}(1-f_{i,\nu^{\prime\prime}=0})\nu_i^{1/2}\Phi(\nu^\prime)
\label{eq:op}
\end{equation}
where,
\begin{equation}
 \Phi(\nu^\prime) = 
\begin{cases}
{1 \over 2}\left[\text{erf}\left({\nu^\prime-\nu_i
      \over \Delta\nu_{i}}\right)-\text{erf}\left({\nu-\nu_i \over
      \Delta\nu_{i}}\right)\right], & \text{for Gaussian profile}\\
\\
{1 \over \pi}\left[\arctan\left({\nu^\prime-\nu_i
      \over \Gamma_i}\right)-\arctan\left({\nu-\nu_i \over
      \Gamma_i}\right)\right],& \text{for Lorenzian profile}
\end{cases}
\label{eq:prof}
\end{equation}
and $\text{erf}(x)$ is the error function.  In Figure~\ref{fig:tau},
we show the H$_2$ {\em total} optical depth (\ie, up to the maximum
$z_{em}$ visible to the observer) of the IGM in the Lyman-Werner bands
at redshifts $z=30$ (solid line) and $z=15$ (dashed line). The opacity
at energies less than Ly$\beta$ ($h\nu < 12.09$ eV) is produced by the
H$_2$ Lyman lines.
At energies higher than Ly$\beta$, the H$_2$ Werner lines are also
important.  The maximum opacity that we find is $\tau \sim 2$, about 6
times higher than that found by \cite{HaimanAR:00}. The background
flux is thus reduced by an order of magnitude if we assume an average
molecular fraction $x_{H_2}=2 \times 10^{-6}$.

\def\capfl{{\em Total} H$_2$ optical depth of the IGM in the
 Lyman-Werner
 bands. The solid line is for an observer at redshift
  $z_{ob}=30$ and the dashed line at $z_{ob}=15$. We assume a constant
  $x_{H_2}=2 \times 10^{-6}$.}
\placefig{
\begin{figure*}[htp]
\plotone{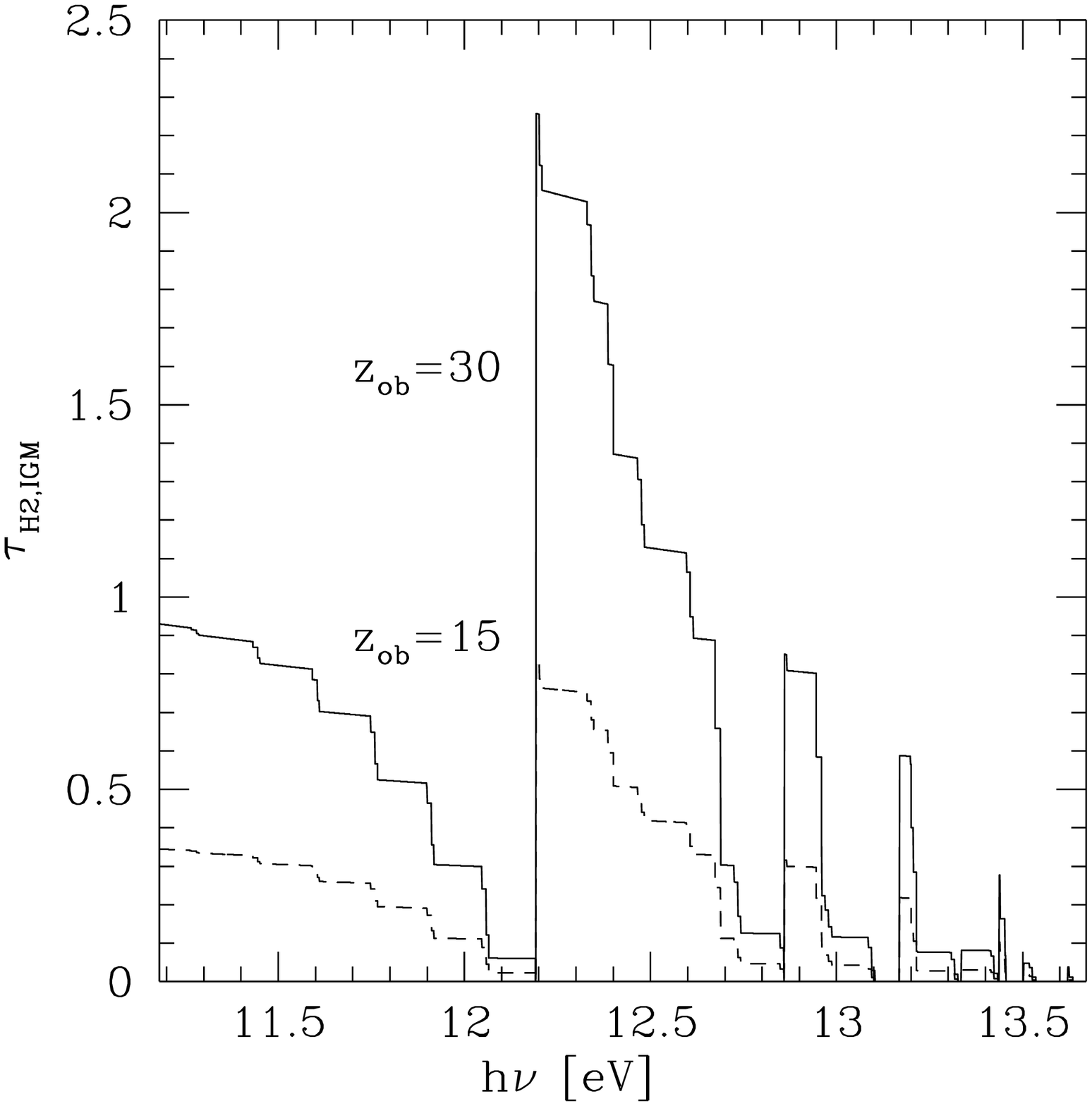}
\caption{\label{fig:tau}\capfl}
\end{figure*}
} 
\def\capfm{In the plot we show the function
[$\text{erf}(x^\prime)-\text{erf}(x)$], where $x^\prime=(\nu^\prime-\nu_i)/
\Delta\nu_{i}$ and $x=(\nu-\nu_i)/ \Delta\nu_{i}$. According to
equation~(\ref{eq:op}), this function is proportional to the optical
depth of an H$_2$ line through a gas of constant density and molecular
fraction. When the source redshift $z_{em}$ increases with
respect to the observer redshift $z_{ob}$, the profile becomes wider,
and the optical depth tends toward the asymptotic value $\tau_{max}$
in equation~(\ref{eq:taumax}). The profiles are shown at
$\log[(z_{em}-z_{ob})/(1+z_{ob})]=-3.5, -4, -4.5, ..., -7$. The gas temperature is either $T=10^4$ K
(solid lines) and $T=10$ K (dashed lines). }
\placefig{
\begin{figure*}[htp]
\plotone{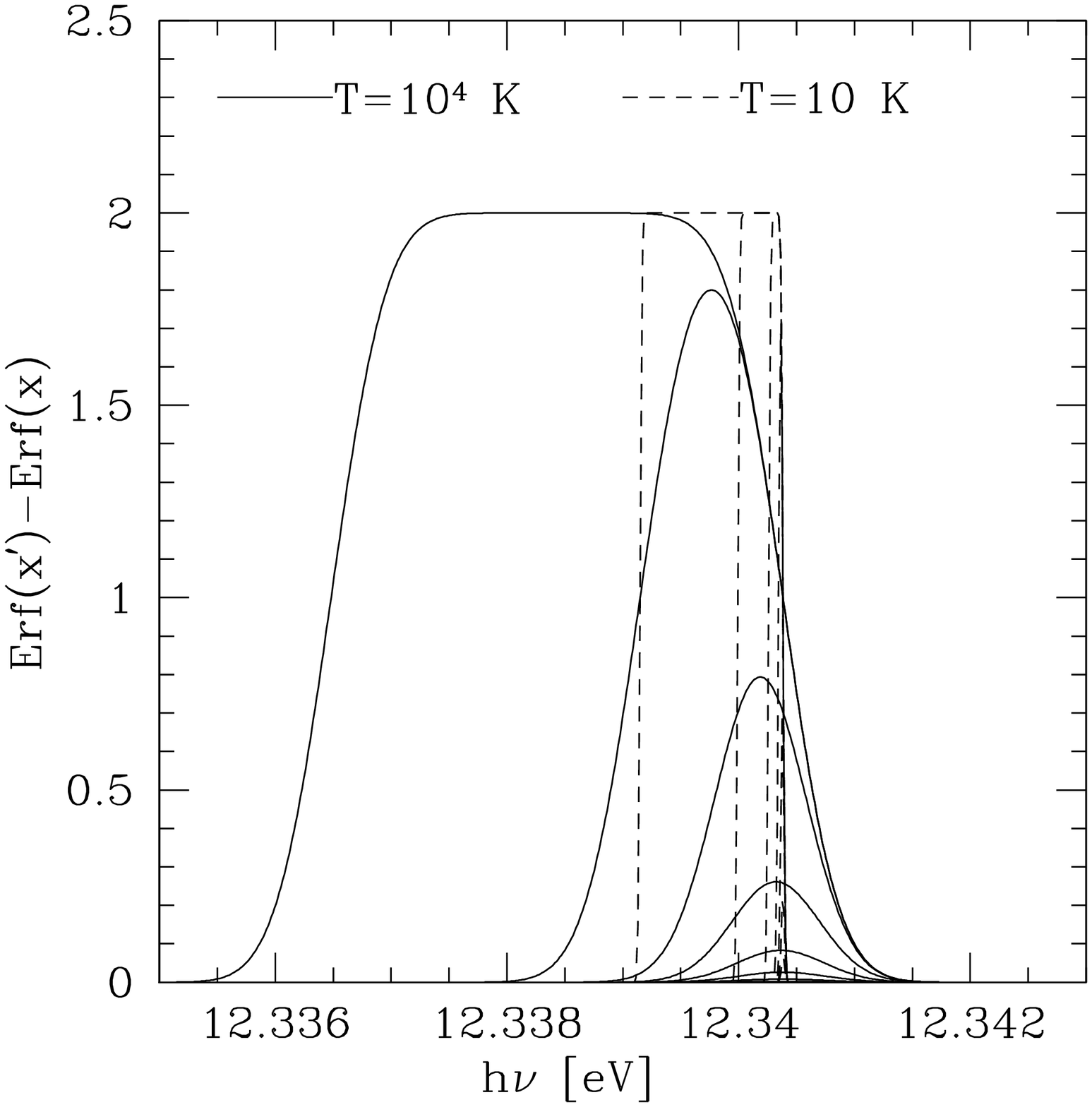}
\caption{\label{fig:oneline}\capfm}
\end{figure*}
}

\def\capfn{Total optical depth, $\langle \tau_{IGM} \rangle$, for
  photo-dissociation of H$_2$ in the IGM as a function of the comoving
  distance from a source at $z_{em}=30$. $\langle \tau_{IGM} \rangle$
  includes the opacity of H$_2$ and H~I lines and is weighted on the
  photo-dissociation rate; see equation~(\ref{eq:k}). The solid line
  is calculated assuming a Voigt profile for the Lyman series hydrogen
  lines, therefore including the effect of the damping wings. The
  dashed line is calculated neglecting H~I damping wings.  The insert
  is a zoom of the inner few Mpc, and the arrow indicates the distance at
  which the absorption lines are redshifted about one line width for a
  gas at $T=10^4$ K.}
\placefig{
\begin{figure*}[htp]
\plotone{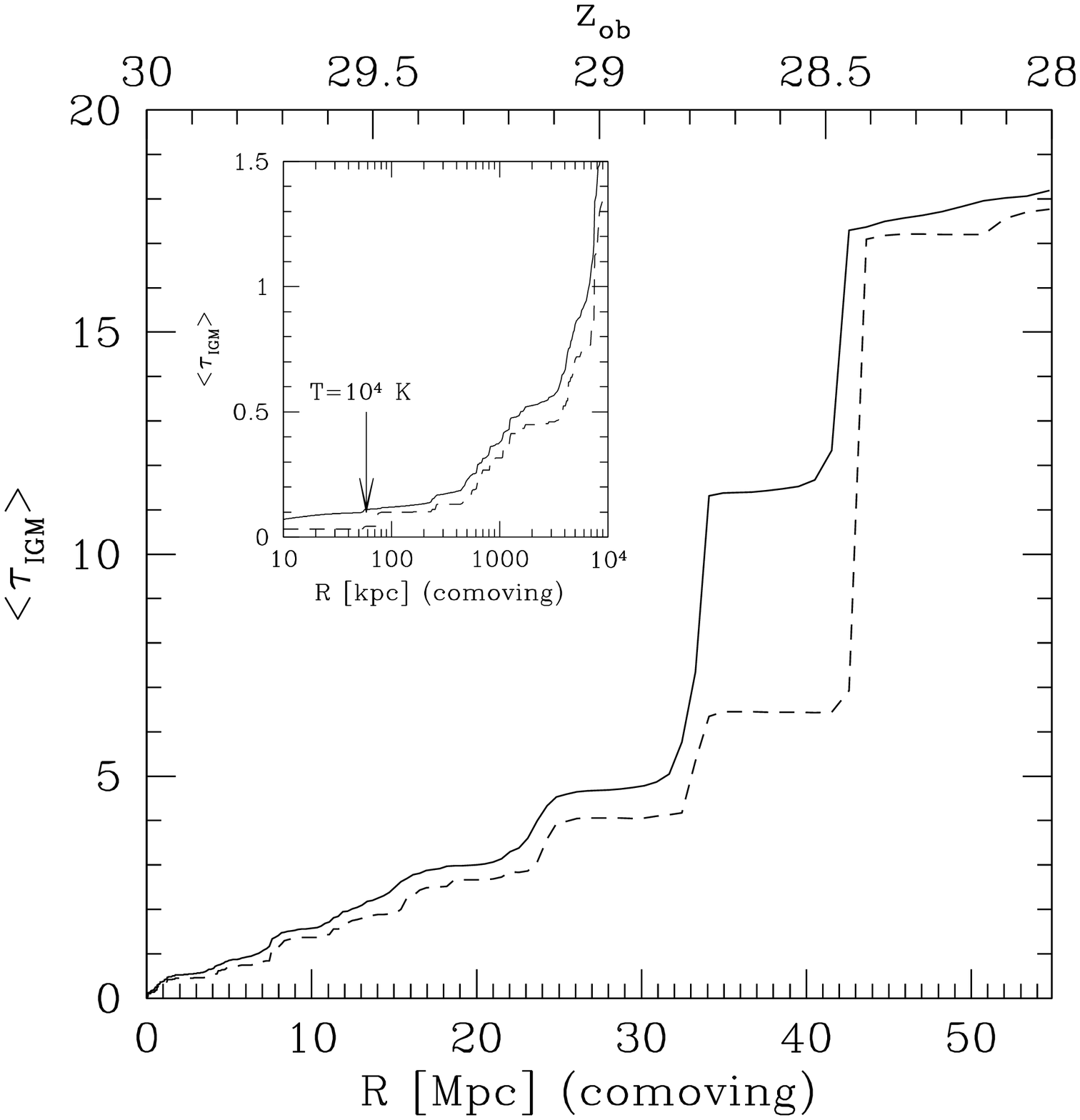}
\caption{\label{fig:k30}\capfn}
\end{figure*}
}

For the Gaussian line profile, the expression inside the square
brackets of equation~(\ref{eq:prof}) tends to the asymptotic value of
2 when $(1+z_{em})/(1+z_{ob})> \Delta\nu_{i}$. The maximum optical
depth of the $i^{th}$ line is therefore
\begin{equation}
\tau_{max}(z_{ob})\simeq x_{H_2}n_H(z=0){c \over H_0}\sigma
f_{osc,i}{(1+z_{ob})^{3/2} \over \nu_i} \simeq 0.1 \left({x_{H_2}
    \over 2 \times 10^{-6}}\right) \left({12~{\rm eV} \over h\nu_i}\right)
\left({f_{osc,i} \over 10^{-2}}\right) \left({1+z_{ob} \over 30}\right)^{3/2}.
\label{eq:taumax}
\end{equation}
In Figure~\ref{fig:oneline}, we show the optical depth through a
constant density and molecular fraction gas for a single line. When
the source redshift $z_{em}$ increases, the profile becomes wider and
the optical depth tends to the asymptotic value $\tau_{max}$ in
equation~(\ref{eq:taumax}).

Finally, we calculate the total optical depth of the IGM as a function
of distance from a single source. We also include in the calculation
the Lyman series lines of H~I and we investigate the importance
of the damping-wings of the heavily saturated lines. The mean optical depth
for photo-dissociation, $\langle \tau_{IGM} \rangle$, can be calculated by
solving the following equations:
\begin{eqnarray}
k_{diss}&=& \text{const} \sum_{i=1}^{76} f_{osc,i}
f_{diss}\int_\nu d\nu^\prime {J_{\nu^\prime} \over h \nu^\prime} ~\phi(\nu^\prime,\nu_i), \\
k_{diss}\exp{(-\langle \tau_{IGM}\rangle)}&=&\text{const} \sum_{i=1}^{76} f_{osc,i}
f_{diss} \int_\nu d\nu^\prime {J_{\nu^\prime} \over h \nu^\prime} ~\phi(\nu^\prime,\nu_i) e^{-\tau_{IGM}(\nu^\prime,z_{em},z_{ob})}
\label{eq:k}
\end{eqnarray}
where $k_{diss}$ is the photo-dissociation rate and where const=$4\pi
x_{H_2}n_H(z_{ob})\sigma$ with $\sigma=(\pi e^2/m_e c)$. Note that
$\tau_{IGM}$ differs from $\tau_\nu$ in equation~(\ref{eq:op}) since
it includes the opacity of both H$_2$ and H~I lines. In
Figure~\ref{fig:k30} we show $\langle \tau_{IGM} \rangle$ as a
function of the comoving distance (and $z_{ob}<z_{em}$) from a source
at $z_{em}=30$. The solid line is calculated assuming a Voigt profile
for the Lyman series hydrogen lines, therefore including the effect of
line saturation. The dashed line is calculated neglecting line
saturation.  The insert is a zoom of the inner few Mpc, and the arrow
indicates the distance at which the absorption lines are redshifted
about one line width for gas at $T=10^4$ K. Both Figure~\ref{fig:k30}
and Figure~\ref{fig:LWspec} show clearly that the opacity at comoving
distances from the source of less than 1 Mpc is produced in the PFR
shell. The IGM opacity starts to be important around 10 Mpc. This
reduces the photo-dissociation radius of the strongest sources at high
redshift.
 
We remind the reader that all the calculations in this section assume
a constant molecular fraction in the IGM of $x_{H_2}=2 \times
10^{-6}$.  In order to calculate a realistic average molecular
fraction in the IGM as a function of the redshift, we need to
calculate the filling factor of the photo-dissociation regions, PFR
and the H$_2$ formed inside relic H~II regions. Only with a 3D
radiative transfer simulation will it be possible to treat
self-consistently the aforementioned feedback effects.

\section{Discussion and Summary \label{sec:sum}}

The results of this work reinforce the possibility that the
population of small mass primordial galaxies (Pop~III) could exist,
without being immediately suppressed by radiative feedback. A
quantitative answer to this question has to be given by
high-resolution cosmological simulations. Until the build-up of the
dissociating background, local feedback effects regulate star
formation.  When the star formation is suppressed in a halo, it is
reasonable to expect that H~II regions start to recombine. The relic
H~II region, as shown in \S~\ref{sec:res}, produces new molecular
hydrogen and perhaps a second burst of star formation. This is
especially effective at high redshifts where the density of the IGM
and protogalaxies is higher and the physical scales are smaller. Thus,
at these redshifts, it is easier to believe that star formation is
bursting rather than continuous. The picture that could emerge from
numerical simulations is that H~II regions and photo-dissociation regions in
the IGM are short-lived instead of continuously expanding, as
in the reionization simulations of \cite{Gnedin:00}. This will have
important effects on the distribution of the metals and the chemical
evolution of galaxies.

In summary, our key results are:

\noindent
(1) We have found a new positive feedback effect on the formation of
H$_2$.  Each source of radiation produces a shell (PFR: {\em positive
  feedback region}) of H$_2$ in front of the H~II region with a
thickness of several kpc and peak abundance $x_{H_2} \sim 10^{-4}$.
The H$_2$ column density of the PFR is typically
$N_F(H_2) \sim 10^{14}-10^{15}$ cm\mm. Fossil H~II regions, if they exist,
are an important mechanism of H$_2$ production, both in the IGM and
inside protogalaxies.

\noindent
(2) The PFR can be optically thick in the lines of the H$_2$
Lyman-Werner bands. The implication is twofold: (i) the
photo-dissociation region around each single source is about 1.5 times
smaller than in the optically thin case; (ii) the H$_2$ optical depth
of the IGM increases. In a later paper, we will calculate the IGM
optical depth and the effect on the intensity of the
photo-dissociating background by means of cosmological simulations.

\noindent
(3) We provide analytical formulae that fit the simulation results,
in order to make a parametric study of the importance of positive
feedback as a function of redshift. The most important parameters
are \fesc and the SED
of the sources (Pop~III, Pop~II or AGN). If \fesc is not extremely
small, PFRs have important effects on the opacity of the IGM to the
H$_2$ photo-dissociating background and the size of photo-dissociation
regions.

\noindent
(4) The background opacity of the IGM in the H$_2$ Lyman-Werner bands
is about unity if $x_{H_2}=2 \times 10^{-6}$.  Therefore, if the relic
molecular hydrogen is not immediately destroyed, it can decrease
the photo-dissociating background flux by about an order of magnitude.

The aforementioned results are the foundations on which we will
construct a cosmological simulation in which these effects are treated
self-consistently.

\acknowledgements 
This work was supported by the Theoretical
Astrophysics program at the University of Colorado (NSF grant AST
96-17073 and NASA grant NAG5-7262). We thank Mark Giroux for a
critical review of the manuscript and Jason Tumlinson for giving us
his Pop~III SED.
\bibliographystyle{/users1/casa/ricotti/Latex/TeX/apj}
\bibliography{/users1/casa/ricotti/Latex/TeX/archive}

\begin{thebibliography}{}

\bibitem[\protect\citeauthoryear{{Abel} et~al.}{{Abel}
  et~al.}{1998}]{AbelAnninos:98}
{Abel}, T., {Anninos}, P., {Norman}, M.~L.,  \& {Zhang}, Y. 1998, \apj, 508,
  518

\bibitem[\protect\citeauthoryear{{Abel} et~al.}{{Abel} et~al.}{1997}]{Abel:97}
{Abel}, T., {Anninos}, P., {Zhang}, Y.,  \& {Norman}, M.~L. 1997, New
  Astronomy, 2, 181

\bibitem[\protect\citeauthoryear{{Abel}, {Bryan}, \& {Norman}}{{Abel}
  et~al.}{2000}]{Abel:00}
{Abel}, T., {Bryan}, G.~L.,  \& {Norman}, M.~L. 2000, \apj, 540, 39

\bibitem[\protect\citeauthoryear{{Black} \& {Dalgarno}}{{Black} \&
  {Dalgarno}}{1976}]{BlackD:76}
{Black}, J.~H.,  \& {Dalgarno}, A. 1976, \apj, 203, 132

\bibitem[\protect\citeauthoryear{{Bromm}, {Coppi}, \& {Larson}}{{Bromm}
  et~al.}{1999}]{BrommCL:99}
{Bromm}, V., {Coppi}, P.~S.,  \& {Larson}, R.~B. 1999, \apjl, 527, L5

\bibitem[\protect\citeauthoryear{{Burles} \& {Tytler}}{{Burles} \&
  {Tytler}}{1998}]{Burles:98}
{Burles}, S.,  \& {Tytler}, D. 1998, \apj, 499, 699

\bibitem[\protect\citeauthoryear{{Ciardi}, {Ferrara}, \& {Abel}}{{Ciardi}
  et~al.}{2000}]{CiardiFA:00}
{Ciardi}, B., {Ferrara}, A.,  \& {Abel}, T. 2000, \apj, 533, 594

\bibitem[\protect\citeauthoryear{{Ciardi} et~al.}{{Ciardi}
  et~al.}{2000}]{CiardiF:00}
{Ciardi}, B., {Ferrara}, A., {Governato}, F.,  \& {Jenkins}, A. 2000, \mnras,
  314, 611

\bibitem[\protect\citeauthoryear{{Croft} et~al.}{{Croft}
  et~al.}{1999}]{Croft:99}
{Croft}, R. A.~C., {Weinberg}, D.~H., {Pettini}, M., {Hernquist}, L.,  \&
  {Katz}, N. 1999, \apj, 520, 1

\bibitem[\protect\citeauthoryear{{de Bernardis} et~al.}{{de Bernardis}
  et~al.}{2000}]{deBernardis:00}
{de Bernardis}, P., et~al. 2000, \nat, 404, 955

\bibitem[\protect\citeauthoryear{{Donahue} \& {Shull}}{{Donahue} \&
  {Shull}}{1987}]{Donahue:87}
{Donahue}, M.,  \& {Shull}, J.~M. 1987, \apj, 323, L13

\bibitem[\protect\citeauthoryear{{Donahue} \& {Shull}}{{Donahue} \&
  {Shull}}{1991}]{Donahue:91}
{Donahue}, M.,  \& {Shull}, J.~M. 1991, \apj, 383, 511

\bibitem[\protect\citeauthoryear{{Dove}, {Shull}, \& {Ferrara}}{{Dove}
  et~al.}{2000}]{Dove:00}
{Dove}, J.~B., {Shull}, J.~M.,  \& {Ferrara}, A. 2000, \apj, 531, 846

\bibitem[\protect\citeauthoryear{{Ferrara}}{{Ferrara}}{1998}]{Ferrara:98}
{Ferrara}, A. 1998, \apjl, 499, L17

\bibitem[\protect\citeauthoryear{{Fuller} \& {Couchman}}{{Fuller} \&
  {Couchman}}{2000}]{Fuller:00}
{Fuller}, T.~M.,  \& {Couchman}, H. M.~P. 2000, \apj, 544, 6

\bibitem[\protect\citeauthoryear{{Galli} \& {Palla}}{{Galli} \&
  {Palla}}{1998}]{Galli:98}
{Galli}, D.,  \& {Palla}, F. 1998, \aap, 335, 403

\bibitem[\protect\citeauthoryear{{Garnavich} et~al.}{{Garnavich}
  et~al.}{1998}]{Garnavich:98}
{Garnavich}, P.~M., et~al. 1998, \apj, 509, 74

\bibitem[\protect\citeauthoryear{{Glover} \& {Brand}}{{Glover} \&
  {Brand}}{2001}]{Glover:00}
{Glover}, S. C.~O.,  \& {Brand}, P. W. J.~L. 2001, \mnras, 321, 385

\bibitem[\protect\citeauthoryear{{Gnedin}}{{Gnedin}}{2000}]{Gnedin:00}
{Gnedin}, N.~Y. 2000, \apj, 535, 530

\bibitem[\protect\citeauthoryear{{Gnedin} \& {Gnedin}}{{Gnedin} \&
  {Gnedin}}{1998}]{GnedinG:98}
{Gnedin}, N.~Y.,  \& {Gnedin}, O.~Y. 1998, \apj, 509, 11

\bibitem[\protect\citeauthoryear{{Haiman}, {Abel}, \& {Rees}}{{Haiman}
  et~al.}{2000}]{HaimanAR:00}
{Haiman}, Z., {Abel}, T.,  \& {Rees}, M.~J. 2000, \apj, 534, 11

\bibitem[\protect\citeauthoryear{{Haiman} \& {Loeb}}{{Haiman} \&
  {Loeb}}{1997}]{Haiman:97}
{Haiman}, Z.,  \& {Loeb}, A. 1997, \apj, 483, 21

\bibitem[\protect\citeauthoryear{{Haiman}, {Rees}, \& {Loeb}}{{Haiman}
  et~al.}{1996}]{HaimanRL:96}
{Haiman}, Z., {Rees}, M.~J.,  \& {Loeb}, A. 1996, \apj, 467, 522

\bibitem[\protect\citeauthoryear{{Hamilton} \& {Tegmark}}{{Hamilton} \&
  {Tegmark}}{2000}]{Hamilton:00}
{Hamilton}, A. J.~S.,  \& {Tegmark}, M. 2000, submitted to \mnras\
  (astro-ph/0008392)

\bibitem[\protect\citeauthoryear{{Hollenbach} \& {McKee}}{{Hollenbach} \&
  {McKee}}{1979}]{Hollenbach:79}
{Hollenbach}, D.,  \& {McKee}, C.~F. 1979, \apjs, 41, 555

\bibitem[\protect\citeauthoryear{{Hui} \& {Gnedin}}{{Hui} \&
  {Gnedin}}{1997}]{Hui:97}
{Hui}, L.,  \& {Gnedin}, N.~Y. 1997, \mnras, 292, 27

\bibitem[\protect\citeauthoryear{{Lange} et~al.}{{Lange}
  et~al.}{2001}]{Lange:00}
{Lange}, A., et~al. 2001, Phys.~Rev.~D, 63, 042001

\bibitem[\protect\citeauthoryear{{Leitherer} et~al.}{{Leitherer}
  et~al.}{1995}]{Leitherer:95}
{Leitherer}, C., {Ferguson}, H.~C., {Heckman}, T.~M.,  \& {Lowenthal}, J.~D.
  1995, \apjl, 454, L19

\bibitem[\protect\citeauthoryear{{Lepp} \& {Shull}}{{Lepp} \&
  {Shull}}{1984}]{Lepp:84}
{Lepp}, S.,  \& {Shull}, J.~M. 1984, \apj, 280, 465

\bibitem[\protect\citeauthoryear{{Machacek}, {Bryan}, \& {Abel}}{{Machacek}
  et~al.}{2001}]{Machacek:00}
{Machacek}, M.~E., {Bryan}, G.~L.,  \& {Abel}, T. 2001, \apj, 548, 509

\bibitem[\protect\citeauthoryear{{Martin}, {Schwarz}, \& {Mandy}}{{Martin}
  et~al.}{1996}]{Martin:96}
{Martin}, P.~G., {Schwarz}, D.~H.,  \& {Mandy}, M.~E. 1996, \apj, 461, 265

\bibitem[\protect\citeauthoryear{{McCray} \& {Kafatos}}{{McCray} \&
  {Kafatos}}{1987}]{McCray:87}
{McCray}, R.,  \& {Kafatos}, M. 1987, \apj, 317, 190

\bibitem[\protect\citeauthoryear{{McDonald} et~al.}{{McDonald}
  et~al.}{2000}]{McDonald:00}
{McDonald}, P., {Miralda-Escud{\'e}}, J., {Rauch}, M., {Sargent}, W. L.~W.,
  {Barlow}, T.~A., {Cen}, R.,  \& {Ostriker}, J.~P. 2000, \apj, 543, 1

\bibitem[\protect\citeauthoryear{{Nakamura} \& {Umemura}}{{Nakamura} \&
  {Umemura}}{1999}]{Nakamura:99}
{Nakamura}, F.,  \& {Umemura}, M. 1999, \apj, 515, 239

\bibitem[\protect\citeauthoryear{{Navarro}, {Frenk}, \& {White}}{{Navarro}
  et~al.}{1997}]{Navarro:97}
{Navarro}, J.~F., {Frenk}, C.~S.,  \& {White}, S. D.~M. 1997, \apj, 490, 493

\bibitem[\protect\citeauthoryear{{Oh}}{{Oh}}{2001}]{Oh:00}
{Oh}, S.~P. 2001, \apj, 553, 499

\bibitem[\protect\citeauthoryear{{Omukai} \& {Nishi}}{{Omukai} \&
  {Nishi}}{1998}]{OmukaiN:98}
{Omukai}, K.,  \& {Nishi}, R. 1998, \apj, 508, 141

\bibitem[\protect\citeauthoryear{{Omukai} \& {Nishi}}{{Omukai} \&
  {Nishi}}{1999}]{OmukaiN:99}
{Omukai}, K.,  \& {Nishi}, R. 1999, \apj, 518, 64

\bibitem[\protect\citeauthoryear{{Perlmutter} et~al.}{{Perlmutter}
  et~al.}{1998}]{Perlmutter:98}
{Perlmutter}, S., et~al. 1998, \nat, 391, 51

\bibitem[\protect\citeauthoryear{{Ricotti}, {Gnedin}, \& {Shull}}{{Ricotti}
  et~al.}{2001}]{RicottiGS:00}
{Ricotti}, M., {Gnedin}, N.~Y.,  \& {Shull}, J.~M. 2001, in ASP Conf. Ser. 222:
  The Physics of Galaxy Formation

\bibitem[\protect\citeauthoryear{{Ricotti} \& {Shull}}{{Ricotti} \&
  {Shull}}{2000}]{RicottiS:00}
{Ricotti}, M.,  \& {Shull}, J.~M. 2000, \apj, 542, 548

\bibitem[\protect\citeauthoryear{{Riess} et~al.}{{Riess}
  et~al.}{1998}]{Riess:98}
{Riess}, A.~G., et~al. 1998, \aj, 116, 1009

\bibitem[\protect\citeauthoryear{{Shapiro} \& {Giroux}}{{Shapiro} \&
  {Giroux}}{1987}]{Shapiro:87}
{Shapiro}, P.~R.,  \& {Giroux}, M.~L. 1987, \apjl, 321, L107

\bibitem[\protect\citeauthoryear{{Shapiro} \& {Kang}}{{Shapiro} \&
  {Kang}}{1987}]{ShapiroKang:87}
{Shapiro}, P.~R.,  \& {Kang}, H. 1987, \apj, 318, 32

\bibitem[\protect\citeauthoryear{{Shull}}{{Shull}}{1978}]{Shull:78}
{Shull}, J.~M. 1978, \apj, 219, 877

\bibitem[\protect\citeauthoryear{{Stecher} \& {Williams}}{{Stecher} \&
  {Williams}}{1967}]{Stecher:67}
{Stecher}, T.~P.,  \& {Williams}, D.~A. 1967, \apjl, 149, L29

\bibitem[\protect\citeauthoryear{{Tegmark} et~al.}{{Tegmark}
  et~al.}{1997}]{Tegmark:97}
{Tegmark}, M., {Silk}, J., {Rees}, M.~J., {Blanchard}, A., {Abel}, T.,  \&
  {Palla}, F. 1997, \apj, 474, 1

\bibitem[\protect\citeauthoryear{{Tegmark}, {Zaldarriaga}, \&
  {Hamilton}}{{Tegmark} et~al.}{2000}]{Tegmark:00}
{Tegmark}, M., {Zaldarriaga}, M.,  \& {Hamilton}, A. J.~S. 2000, submitted
  (astro-ph/0008167)

\bibitem[\protect\citeauthoryear{{Tumlinson} \& {Shull}}{{Tumlinson} \&
  {Shull}}{2000}]{Tumlinson:00}
{Tumlinson}, J.,  \& {Shull}, J.~M. 2000, \apjl, 528, L65

\bibitem[\protect\citeauthoryear{{Venkatesan}, {Giroux}, \&
  {Shull}}{{Venkatesan} et~al.}{2001}]{Venkatesan:01}
{Venkatesan}, A., {Giroux}, M.~L.,  \& {Shull}, J.~M. 2001, submitted

\end{thebibliography}

\vskip 2truecm

\placefig{\end{document}}

\clearpage

\newcounter{figurecap}
\setcounter{figurecap}{0}

\begin{center}
\bf Figure Captions
\end{center}

\refstepcounter{figurecap}
Fig.\ \thefigurecap---\label{fig:sed1}\capfa

\refstepcounter{figurecap}
Fig.\ \thefigurecap---\label{fig:sed2}\capfb

\refstepcounter{figurecap}
Fig.\ \thefigurecap---\label{fig:1}\capfc

\refstepcounter{figurecap}
Fig.\ \thefigurecap---\label{fig:1b}\capfd

\refstepcounter{figurecap}
Fig.\ \thefigurecap---\label{fig:LWspec}\capfe

\refstepcounter{figurecap}
Fig.\ \thefigurecap---\label{fig:2}\capff

\refstepcounter{figurecap}
Fig.\ \thefigurecap---\label{fig:2a}\capfg

\refstepcounter{figurecap}
Fig.\ \thefigurecap---\label{fig:3}\capfh

\refstepcounter{figurecap}
Fig.\ \thefigurecap---\label{fig:4}\capfi

\refstepcounter{figurecap}
Fig.\ \thefigurecap---\label{fig:tau}\capfl

\refstepcounter{figurecap}
Fig.\ \thefigurecap---\label{fig:k30}\capfm

\refstepcounter{figurecap}
Fig.\ \thefigurecap---\label{fig:oneline}\capfn

\clearpage

\tabone

\end{document}